\documentclass[biomimetics,article,preprints,pdftex,moreauthors]{mdpi}

\firstpage{1} 
\pubvolume{1}
\issuenum{1}
\articlenumber{0}
\pubyear{2024}
\copyrightyear{2024}
\datereceived{ } 
\daterevised{ } %
\dateaccepted{ } 
\datepublished{ } 
\hreflink{https://doi.org/} %

\usepackage{placeins}
\usepackage{mathtools}
\usepackage{graphicx}

\DeclarePairedDelimiter\abs{\lvert}{\rvert}%
\DeclarePairedDelimiter\norm{\lVert}{\rVert}%

\makeatletter
\let\oldabs\abs
\def\abs{\@ifstar{\oldabs}{\oldabs*}}
\let\oldnorm\norm
\def\norm{\@ifstar{\oldnorm}{\oldnorm*}}
\makeatother

\newcommand{\sci}[1] {\times 10^{#1}} %

\newcommand{\degr}{^{\circ}} %

\newcommand{\eg}{e.g.\ }
\newcommand{\aoa}{\text{AoA} }

\newcommand{\re}{Re}

\Title{Numerical Study of Flow Past a Wall-Mounted \\ Dolphin Dorsal Fin at Low Reynolds Numbers}

\TitleCitation{Title}

\Author{Zhonglu Lin $^{1,2,\dagger}$\orcidA{0000-0003-1989-7180}, An-Kang Gao $^{3,\dagger}$\orcidB{0000-0002-9805-1388} and Yu Zhang $^{1,2}$*}

\AuthorNames{Zhonglu Lin, An-Kang Gao and Yu Zhang}

\AuthorCitation{Lin, Z.; Gao, A.-K.; Zhang, Y.}

\address{%
$^{1}$ \quad Key Laboratory of Underwater Acoustic Communication and Marine Information Technology of the Ministry of Education, College of Ocean and Earth Sciences, Xiamen University, Xiamen City, Fujian Province, 361005, China \\
$^{2}$ \quad State Key Laboratory of Marine Environmental Science, College of Ocean and Earth Sciences, Xiamen University, Xiamen City, Fujian Province, 361005, China \\
$^{3}$ \quad Department of Modern Mechanics, University of Science and Technology of China, Hefei 230026, China}

\corres{Correspondence: yuzhang@xmu.edu.cn;}

\firstnote{These authors contributed equally to this work.}

\abstract{
	Dolphin swimming has been a captivating area of study, yet the hydrodynamics of the dorsal fin remain underexplored. In this study, we present three-dimensional simulations of flow around a wall-mounted dolphin dorsal fin, derived from a real dolphin scan. The NEK5000 (spectral element method) is employed with a second-order hex20 mesh to ensure high accuracy and computational efficiency in the simulations.
    A total of 13 cases were simulated, covering angles of attack (AoA) 
    ranging from $0^\circ$ to $60^\circ$ and Reynolds numbers ($\text{Re}$) 
    between 691 and 2000.
    Our results show that both drag and lift increase significantly with the AoA. Almost no vortex is observed at $\aoa = 0\degr$, whereas complex vortex structures emerge for $\aoa \geq 30\degr$, including half-horseshoe, hairpin, arch, and wake vortices.
    This study offers insights that could inform 
    the design of next-generation underwater robots, heat exchangers, 
    and submarine sails.
	}

\keyword{Hydrodynamics 1; Computational Fluid Dynamics 2; Numerical Simulation 3; Dolphin 4; Swimming 5; Bio-locomotion 6;}

\begin{document}

\crefname{figure}{Fig.}{Fig.}
\crefname{equation}{Eq.}{Eq.}

\section{Introduction}

Dolphin hydrodynamics has long been a topic of interest for fluid dynamists, 
robot designers, and biologists \cite{Shoele2015,Weihs2004,Yu2023,Fish1999,Wang2021a,Wang2020,Han2020,Pavlov2021,Fish2008}.
The dolphin's dorsal fin plays a crucial role in its underwater propulsion and manoeuvring, contributing to its adaptability to various aquatic environments.
Studying the hydrodynamic mechanisms of the dorsal fin can provide insights 
into both biological understanding of cetaceans \cite{Nino-Torres2023} and innovative engineering designs for biomimetic underwater robotics \cite{Li2024}, heat exchangers \cite{Bashtani2023}, and submarine sails \cite{Rahmani2024}.

Hydrodynamic studies focusing specifically on the dolphin dorsal fin are relatively rare, and a comprehensive study of the \textit{near-fin flow structure} is yet to be conducted. While previous research has focused on dolphin swimming, with particular emphasis on \textit{dynamic propulsion} \cite{Han2020,Guo2023} and leaping behaviour above water \cite{Bergmann2022}, the current study is more relevant to gliding movements in dolphin-like swimmers \cite{Skrovan1999,Wu2019,Zhang2023}. 
\citeauthor{Pavlov2012} \cite{Pavlov2012} conducted numerical simulations of flow past a static dolphin body with a dorsal fin, demonstrating the minimal hydrodynamic impact of a device attached to the dolphin. 
\citeauthor{Nino-Torres2023} \cite{Nino-Torres2023} discovered high similarities in the dorsal fin shapes of common bottlenose dolphins in the Caribbean Sea, closely resembling the fin geometry used in this study.
\citeauthor{Okamura2021} \cite{Okamura2021} found that the dorsal fin, along with the pectoral fins, enhances stability during straight-line locomotion in small cetaceans, such as dolphins.
Inspired by the streamlined shape of 
the dolphin dorsal fin, \citeauthor{Bashtani2023} \cite{Bashtani2023} designed 
a double-tube heat exchanger with reduced friction loss.
\citeauthor{Han2020} \cite{Han2020} applied the immersed boundary method to study a forward-swimming dolphin 
with an attached dorsal fin and pectoral fins, analysing the surface pressure distribution. \citeauthor{Tanaka2019} \cite{Tanaka2019} measured the time-varying kinematics of a dolphin during burst acceleration, providing a scanned geometry of the dolphin, where the dorsal fin is adapted and applied in the present study. Finally, \citeauthor{Rahmani2024} \cite{Rahmani2024} utilised the dolphin dorsal fin shape as the sail of a submarine. 

Although the \textit{dynamic} whole-body swimming dolphin has already been studied, \eg \cite{Guo2023}, it remains valuable to study the flow past a \textit{static} wall-mounted dolphin dorsal fin. The \textit{static} condition is more relevant for certain applications, such as bionic heat exchangers \cite{Bashtani2023} and submarine sails \cite{Rahmani2024}, where the detailed 3-D vortex structure has yet to be fully explored.
Moreover, in addition to continuous locomotion, dolphins in nature use gliding as an energy-saving strategy during swimming, making the flow past a \textit{static} wall-mounted dorsal fin particularly relevant.
For instance, adult bottlenose dolphins (\textit{Tursiops truncatus}) can glide up to 16 metres during a dive \cite{Skrovan1999}.
Similarly, newborn dolphin calves (0–1 month old) of the same species spend about one-third of their time gliding \cite{Noren2008}, rather than actively propelling themselves.
It is hypothesised that during gliding over such distances, the flow structure around the dolphin is markedly different from that of an actively propelling dolphin.
Additionally, several studies on dolphin-inspired robots have focused specifically on gliding \cite{Wu2019,Zhang2023}.
To the best of our knowledge, the unique vortex structures identified in this study have not been previously documented.
Therefore, despite the pioneering research on dolphin swimming, the authors believe that investigating this \textit{static} problem remains valuable.

In summary, to the best of our knowledge, no comprehensive study has yet focused on the 3-D flow structures near a dolphin's dorsal fin, despite its potential applications \cite{Tawade2020,Bashtani2023,Li2024,Rahmani2024}.
Thus, in this paper, we investigate the hydrodynamics of flow past a wall-mounted dorsal fin, with an emphasis on its vortical structures at Reynolds numbers ranging from $691$ to $2000$, based on the dorsal fin's base chord length, and angles of attack (AoA) from $0\degr$ to $60\degr$.
As live dolphins, dolphin robots, or submarine sails resembling a dorsal fin all glide during forward propulsion, they frequently encounter underwater currents from various angles of attack.
Dolphins have evolved to be highly adapted to the dynamic underwater environment, so it should be reasonable to postulate that their dorsal fins have evolved to provide advantages in such situations, making them worthy of investigation.
Dolphins can glide at an average speed of 1.47 m/s\cite{Skrovan1999}, which corresponds to $\re = 10^5$ in the present study. However, due to constraints in computational resources, the maximum $\re$ simulated in this study is $2000$.
Nevertheless, it is not uncommon to use low Reynolds number ($10^3$) simulations to discuss moderate Reynolds number 
($10^5$) experimental results \cite{Zhong2019}.

\section{Computational Methods}

This section outlines the methodology used in the study.
We employ the open-source software NEK5000 \cite{fischer2007nek5000} to generate accurate spectral element solutions 
for the flow around a wall-mounted dolphin dorsal fin subjected to steady currents, with varying Reynolds numbers ($\re$) and angles of attack (AoA). Additionally, a mesh independence study is conducted to ensure the accuracy of the results. Further validation is provided in \Cref{sec:appendix}.
In this study, the Reynolds number is defined based on the fin's chord length $C$ as $\re = UC/\nu$, where $\nu$ is the kinematic viscosity, and $U$ is the free-stream velocity.

\subsection{NEK5000}
NEK5000 \cite{fischer2007nek5000}, which is based on a spectral element method, is an open-source package developed at Argonne National Laboratory. It is known for its scalability, efficiency and capacity to handle unstructured mesh, which is essential for the curved surface of a dolphin dorsal fin.
In NEK5000, the governing equation of incompressible Navier-Stokes equations \cite{quartapelle2013numerical} with constant fluid properties can written as:
\begin{equation}
	\frac{\partial \mathbf{u}}{\partial t} = -\mathbf{u} \cdot \nabla \mathbf{u} - \nabla p + \frac{1}{Re} \nabla^2 \mathbf{u} + \mathbf{f}, \quad \nabla \cdot \mathbf{u} = 0,
\end{equation}
where $\mathbf{u} = \{u, v, w\}^\mathrm{T}$ is the velocity vector, $p$ is the pressure, and $\mathbf{f}$ is the body-force term. 
NEK5000 is particularly advantageous for handling complex vortex structures and flow fields while maintaining high accuracy. It has also been successfully applied to studies involving accelerating swimmers \cite{Abouhussein2023a}. In the Spectral Element Method (SEM)\cite{Patera1984,Karniadakis2007}, the physical domain is divided into spectral elements. The local approximation of the flow field is then represented as a sum of Lagrange interpolants, which are defined by an orthogonal basis of Legendre polynomials up to degree \(N\). The same polynomial order \(N\) is applied in all spatial directions, with a \(P_N - P_N\) spatial discretisation employed \cite{fischer2007nek5000}. In most simulations, \(N=3\) was used. The equations are advanced in time with a second-order conditionally stable backward differentiation and extrapolation scheme (BDF2/EXT2)\cite{fischer2003implementation,SalehRezaeiravesh2021}, using an implicit approach for the diffusion term and an explicit approach for the advection term.
Nek5000 is highly parallelised and scalable, capable of running on thousands of threads\cite{Tufo2001}. The results presented in this study were obtained using 64 to 128 processors.

\subsection{Problem Setup}
\begin{figure}
	\centering
	\includegraphics[width=1\linewidth]{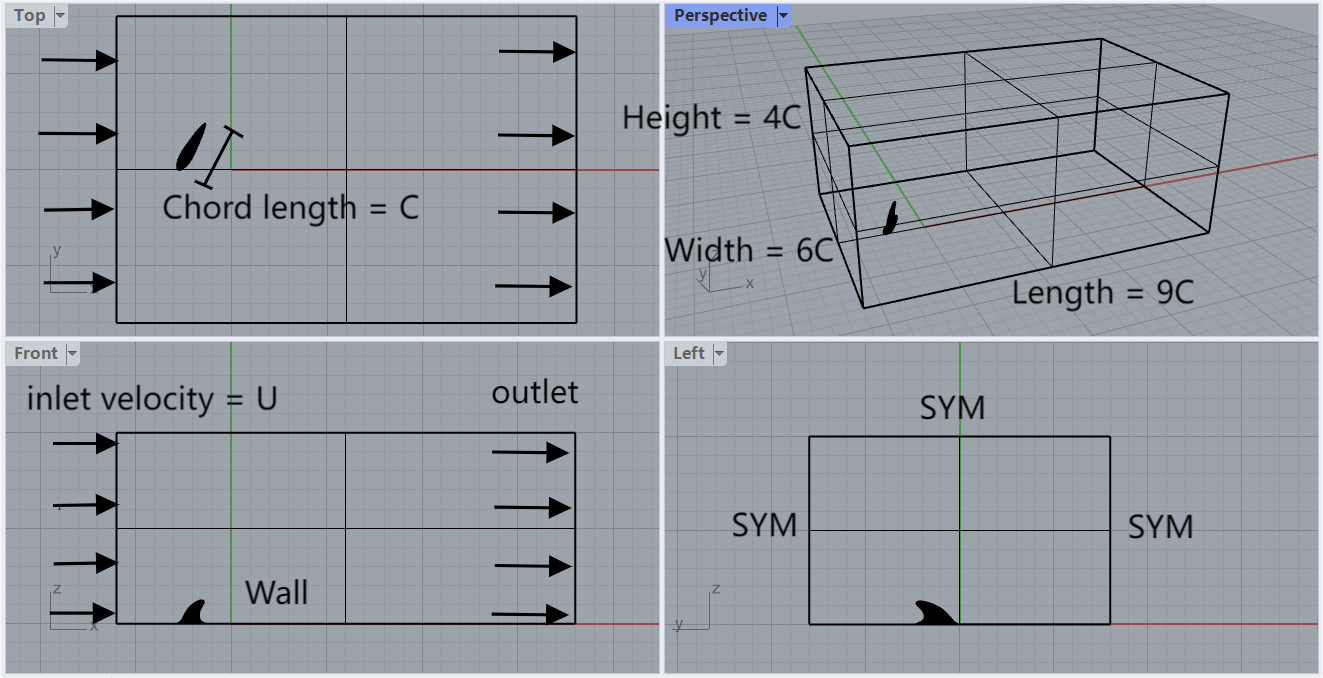}
	\caption{Schematic show of the problem setup, where the "SYM" is the symmetric boundary condition and the "Wall" is the non-slip boundary. "C" is the characteristic length.}
	\label{fig:geobox}
\end{figure}

\begin{figure}
	\centering
	\includegraphics[width=1\linewidth]{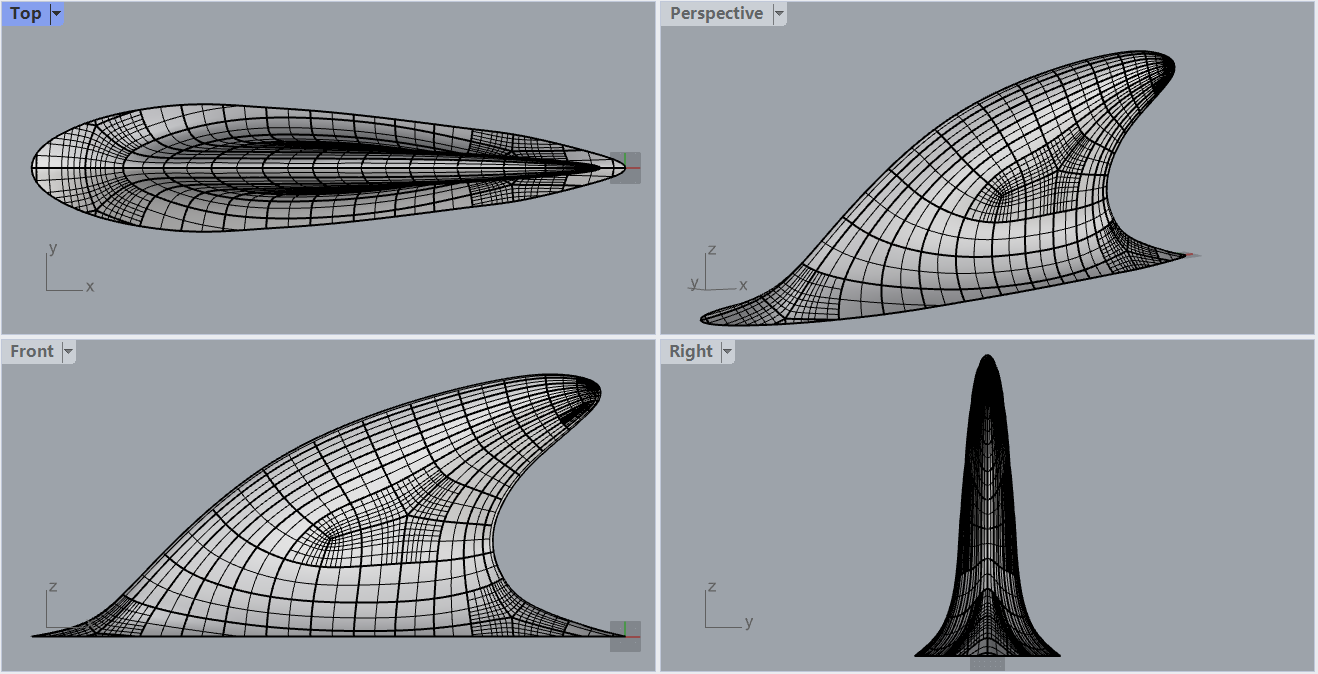}
	\caption{Close-up view of the mesh on the dolphin dorsal fin, adapted from \citeauthor{Tanaka2019} \cite{Tanaka2019}}
	\label{fig:geofin}
\end{figure}

This section outlines the setup of the representative problem. As shown in \Cref{fig:geobox}, within a rectangular computational domain, a dolphin dorsal fin is mounted on the bottom wall, subjected to a steady current of incompressible fluid. Both the dorsal fin and the bottom wall are treated as no-slip boundaries. The inlet velocity is uniform at $U$. Symmetry boundary conditions are applied to the side walls and top wall, effectively acting as slip boundaries. The geometry of the dolphin dorsal fin (\Cref{fig:geofin}) is adapted from the scan of a frozen dolphin (\textit{Lagenorhynchus obliquidens}) by \citeauthor{Tanaka2019} \cite{Tanaka2019}.
For boundary conditions, at the inlet, Dirichlet conditions are prescribed as $\{u, v, w\}^\mathrm{T} = \{ U, 0, 0\}^\mathrm{T}$. For non-stress formulation, the outlet boundary condition can be written as $[-p \mathbf{I} + \nu (\nabla \mathbf{u})] \cdot \hat{\mathbf{e}}_n = 0
$. The symmetric boundary ("SYM" in \Cref{fig:geobox}) is prescribed as $\mathbf{u} \cdot \hat{\mathbf{e}}_n = 0$, $(\nabla \mathbf{u} \cdot \hat{\mathbf{e}}_t) \cdot \hat{\mathbf{e}}_n = 0$, $(\nabla \mathbf{u} \cdot \hat{\mathbf{e}}_b) \cdot \hat{\mathbf{e}}_n = 0$. Here, $\hat{\mathbf{e}}_n$ is the unit normal vector, $\hat{\mathbf{e}}_t$ is the unit tangent vector, and $\hat{\mathbf{e}}_b$ is the unit bitangent vector. The wall boundary condition is $\mathbf{u} = 0$.

\begin{figure}
	\centering
	\includegraphics[width=0.7\linewidth]{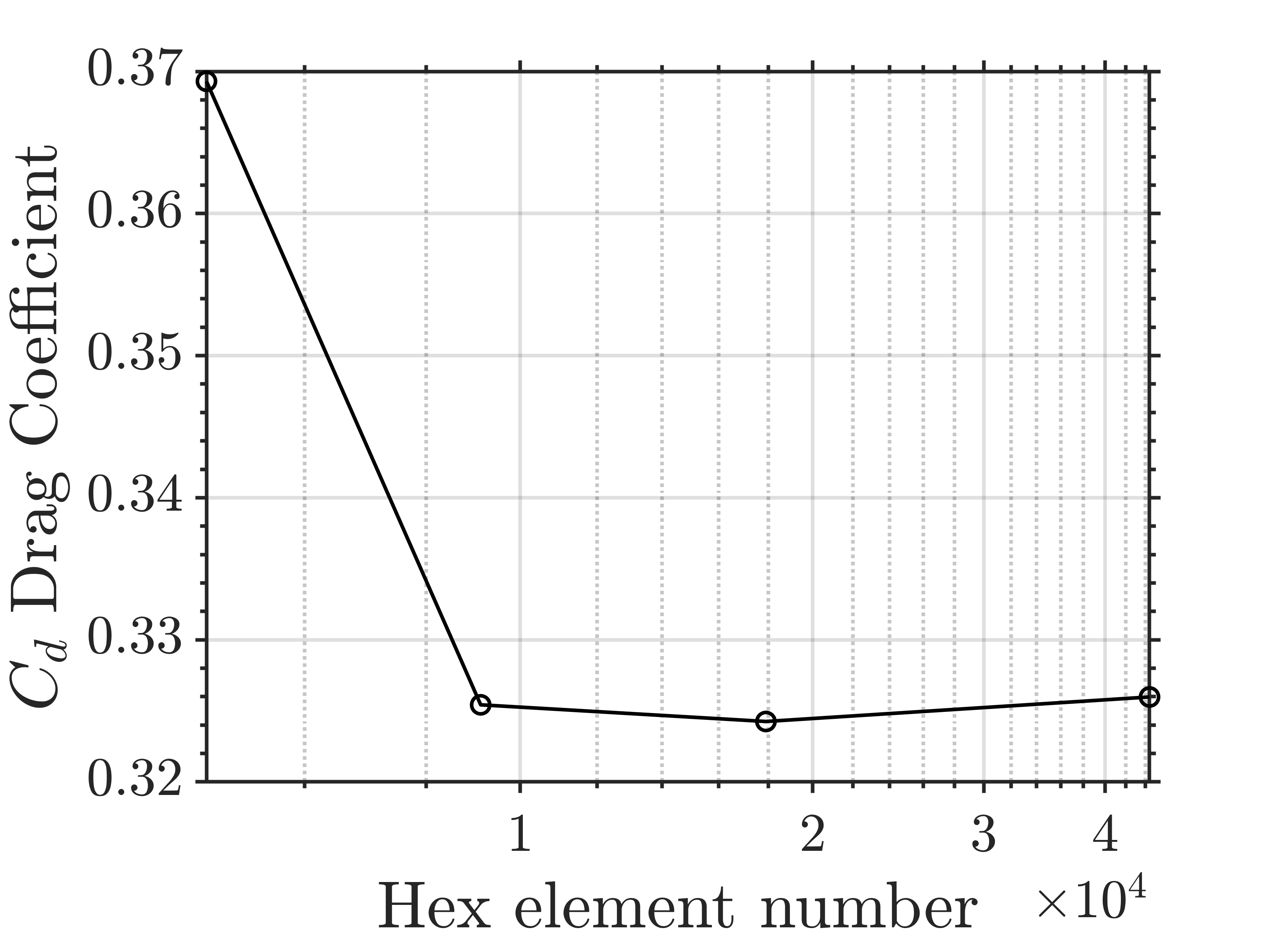}
	\caption{A grid convergence study is conducted at $Re = 691$ and $\aoa = 0\degr$ for four meshes of various mesh densities. The drag coefficient quickly converges at the second mesh. The mesh of 71\,612 hex elements is chosen to conduct the simulations in this paper. The CFL number is controlled at approximately 0.5 for all the simulations. The number of Gauss-Lobatto-Legendre points per element in each spatial direction is fixed at four. %
	}
	\label{fig:meshindp}
\end{figure}

A grid independence study was performed to ensure the selection of an appropriate mesh, as shown in \Cref{fig:meshindp}. The mesh with 71\,612 elements was chosen for the current study.

\begin{figure}
	\centering
	\includegraphics[width=1\linewidth]{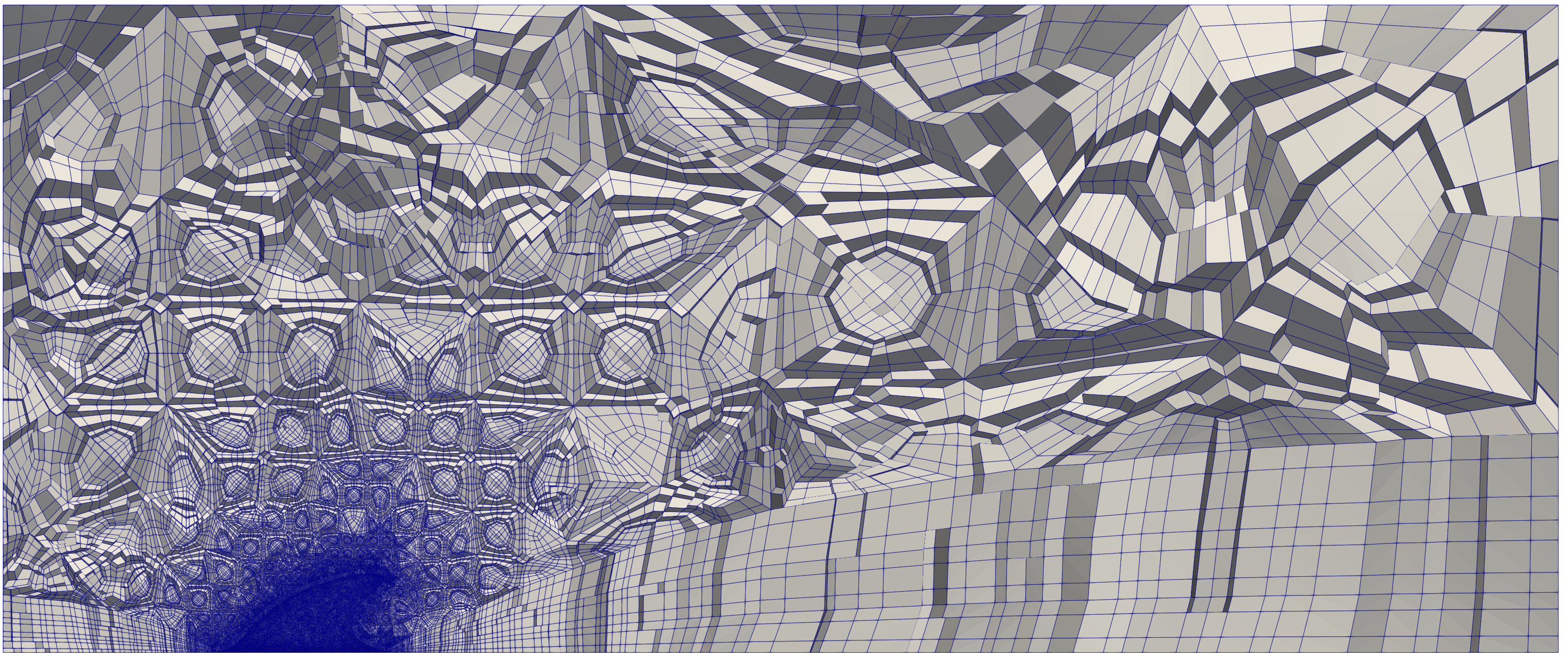}
	\caption{Clipped overview of the selected unstructured pure hex mesh. Non-linear quadratic mesh is applied.}
	\label{fig:meshbig}
\end{figure}

\begin{figure}
	\centering
	\includegraphics[width=1\linewidth]{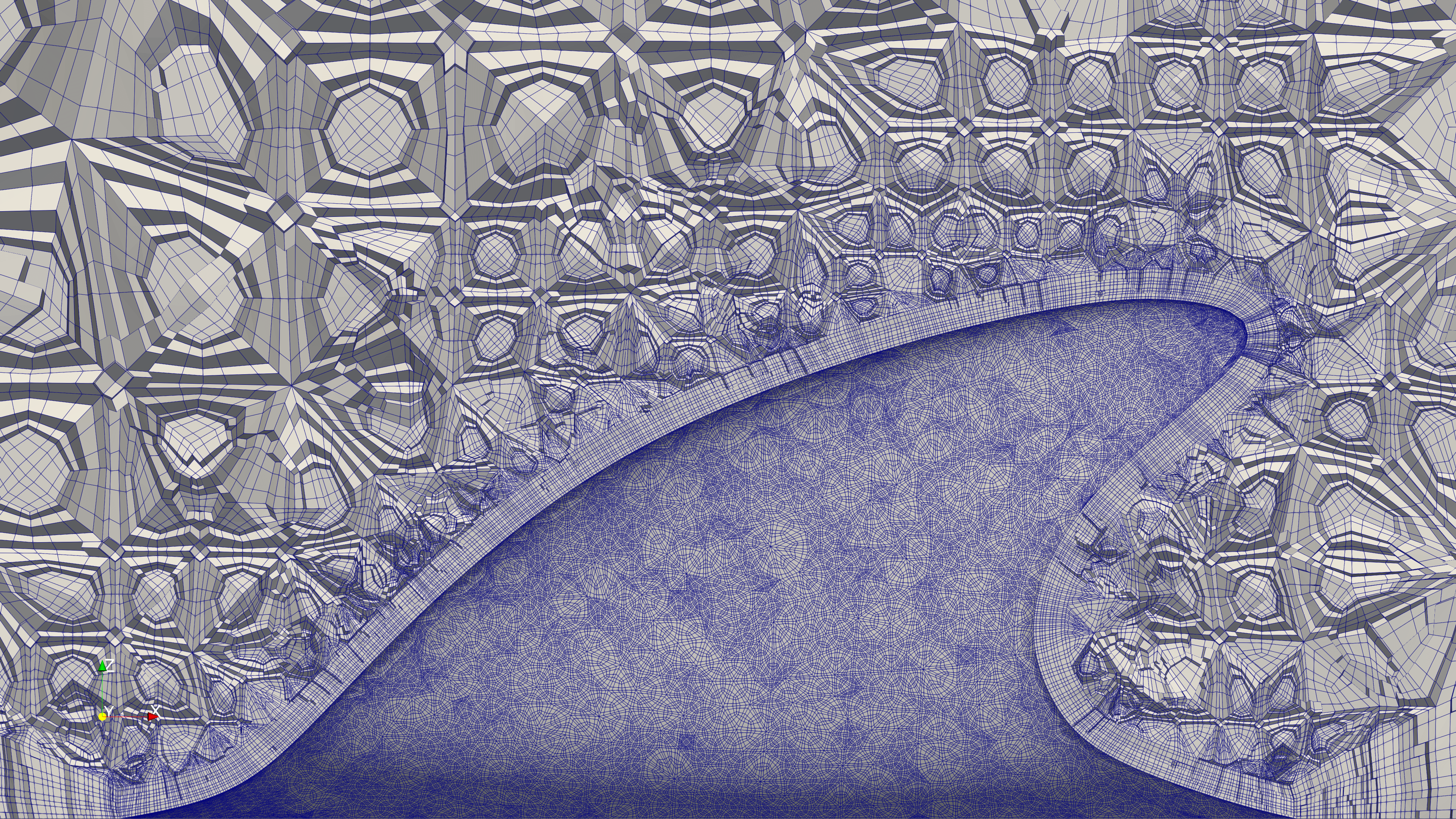}
	\caption{Clipped close-up view of selected mesh near the dolphin dorsal fin}
	\label{fig:meshclose}
\end{figure}

The drag coefficient is defined as $C_d = 2 F_x / (\rho U^2 A)$, where $ F_x $ is the force in the streamwise direction, $\rho$ is the fluid density, and $A$ is the reference area of the fin at $\aoa = 0\degr$ projected onto the side plane.
Global and close-up views of the mesh can be found in \Cref{fig:meshbig} and \Cref{fig:meshclose}.

\section{Discussion of Numerical Results}

Simulations are carried out at $Re = 691-2000$ and $\aoa = 0\degr-60\degr$. We analysed the results by examining the drag coefficient, as well as the isosurfaces of the Q-criterion, pressure, and velocity magnitude. The force coefficient and flow structure show minimal variation with $Re$, but significant changes with $\aoa$. Therefore, our discussion will focus primarily on the variation with $\aoa$.

\begin{figure}
	\centering
	(a)
	\includegraphics[width=0.45\linewidth]{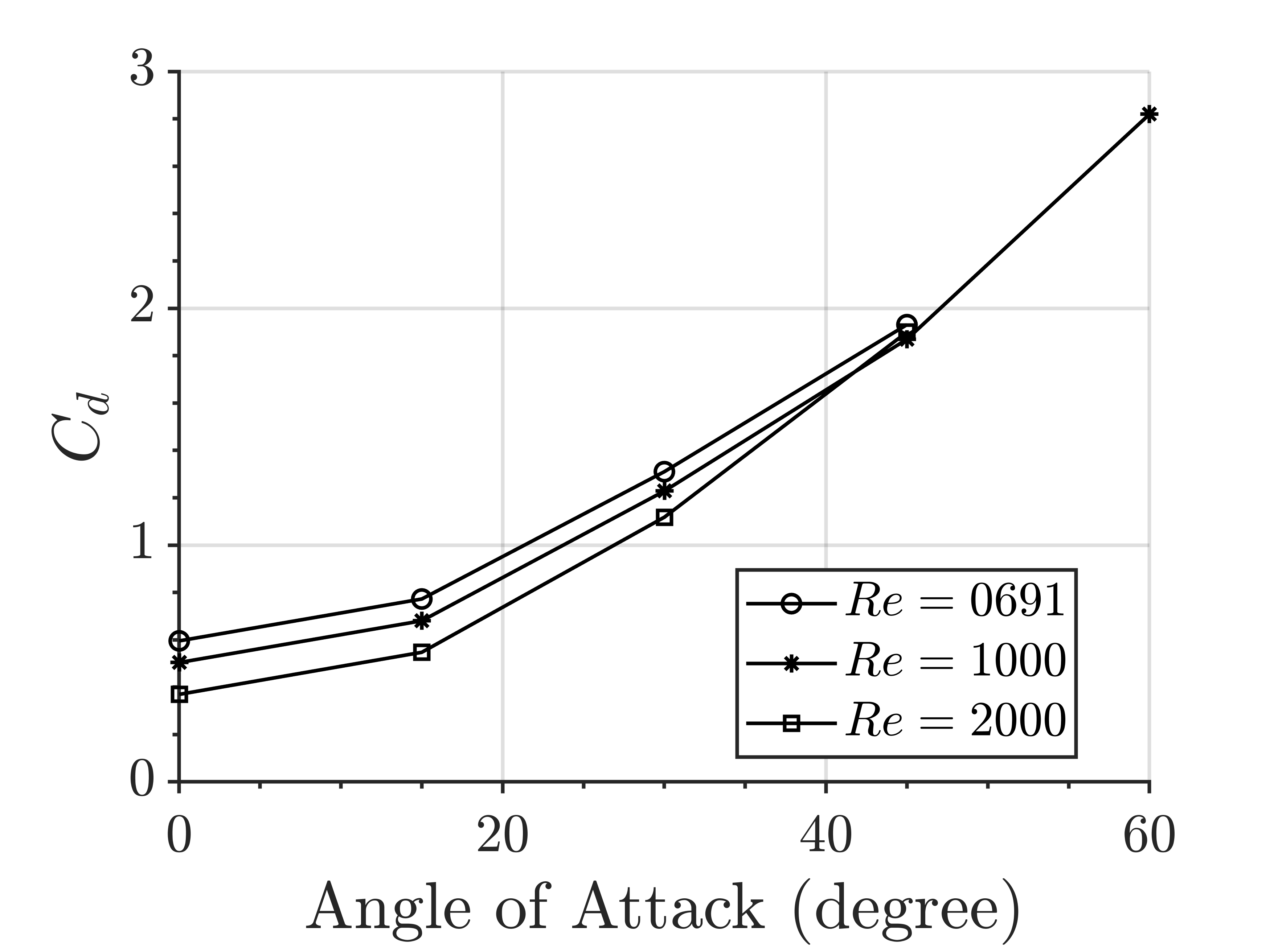}
	(b)
	\includegraphics[width=0.45\linewidth]{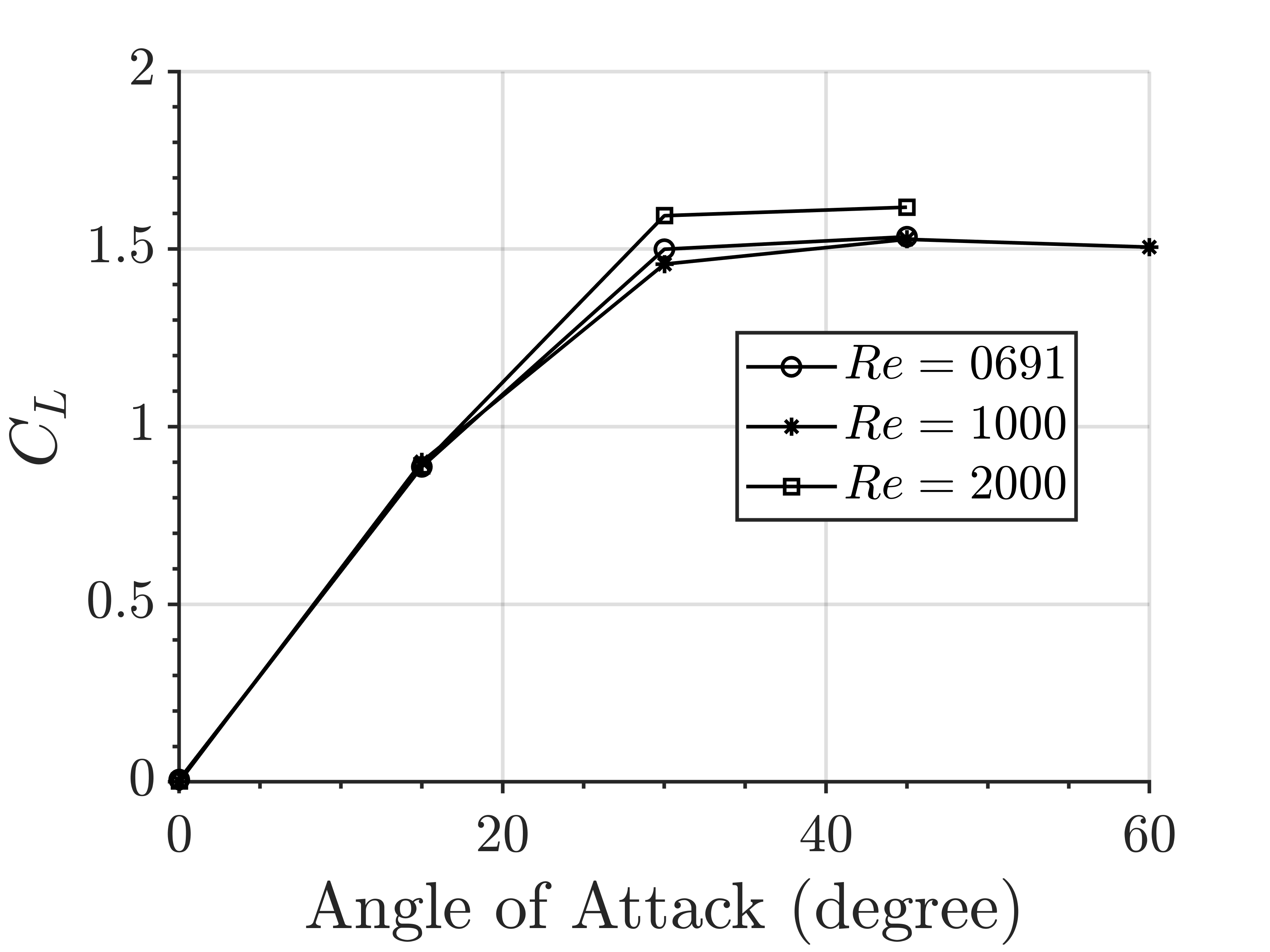}
	\caption{Variation of the drag coefficient $C_d$ and the lift coefficient $C_L$ at $\aoa = 0 - 60\degr$ and $\re = 691 - 2000$. 
	}
	\label{fig:dragcoe}
\end{figure}

At low $\re = 691 - 2000 $, the drag and lift coefficients both increase with the AoA. As shown in \Cref{fig:dragcoe}a, increasing the $\aoa$ from $0\degr$ to $60\degr$ results in more than a fivefold increase in drag ($C_d$). In \Cref{fig:dragcoe}b, the lift coefficient ($C_L$) rises from 0 at $\aoa = 0\degr$ to approximately 1.5 at $\aoa = 30\degr$, before reaching a plateau between $\aoa = 30\degr$ and $60\degr$.
Regarding the effect of the Reynolds number, drag decreases slightly as the Reynolds number increases, while the lift coefficient remains largely unaffected by changes in the Reynolds number.

\FloatBarrier
\subsection{Flow structure by Q-criterion iso-surface}

\begin{figure}
	\centering
	\includegraphics[width=1\linewidth]{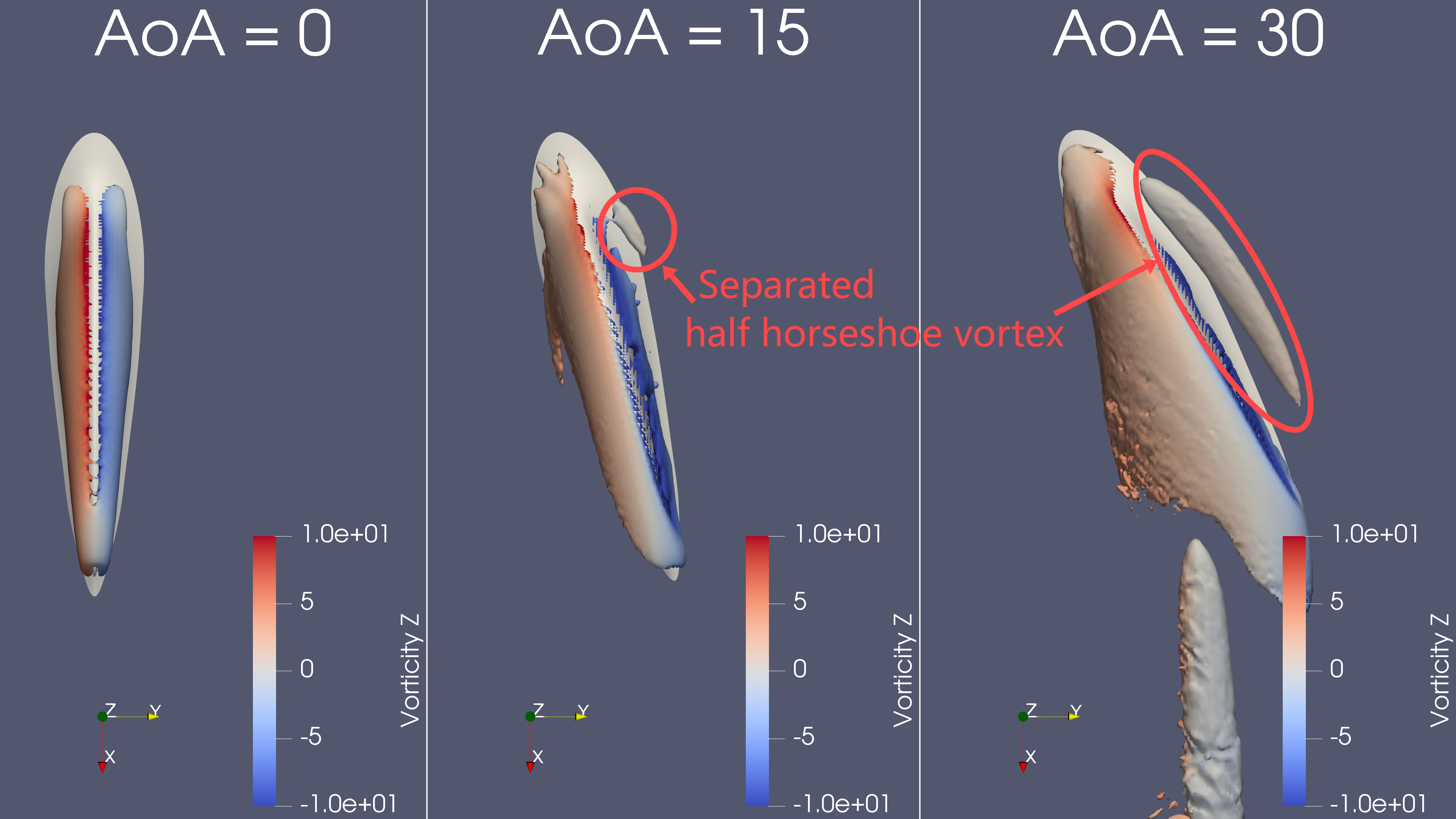}
	\includegraphics[width=1\linewidth]{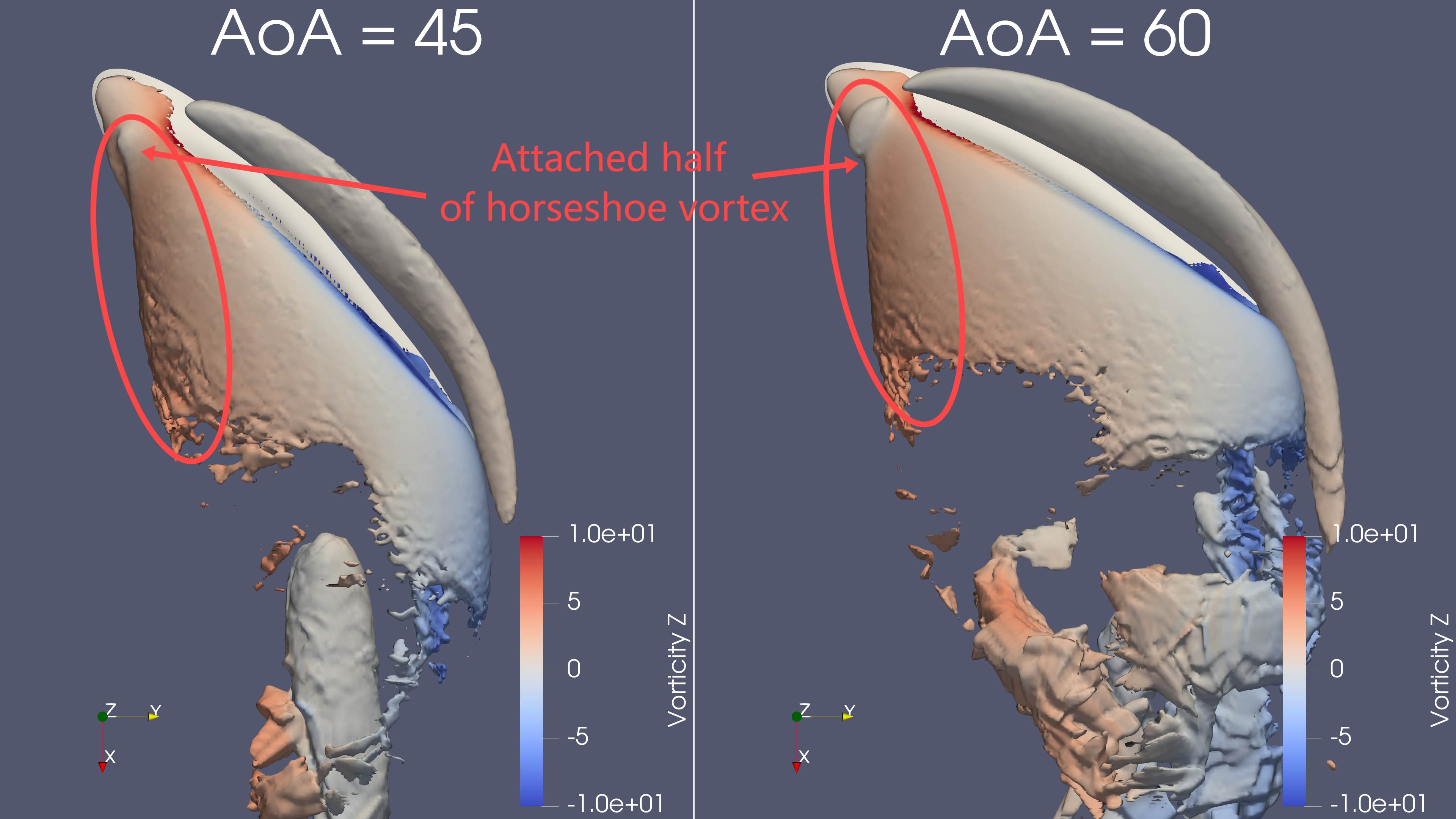}
	\caption{At $\re = 1000$ and $\aoa = 0 - 60\degr$, the isosurface of $Q^* = 7.7 $ is coloured by vorticity in the Z direction viewed from the top.
    The separated half-horseshoe vortex is found near the upstream side, whereas the attached half-horseshoe is upon the downstream surface and near the leading edge.
 }%
	\label{fig:aoas_Re_1000_Q_top}
\end{figure}

\begin{figure}
    \centering
    \includegraphics[width=1\linewidth]{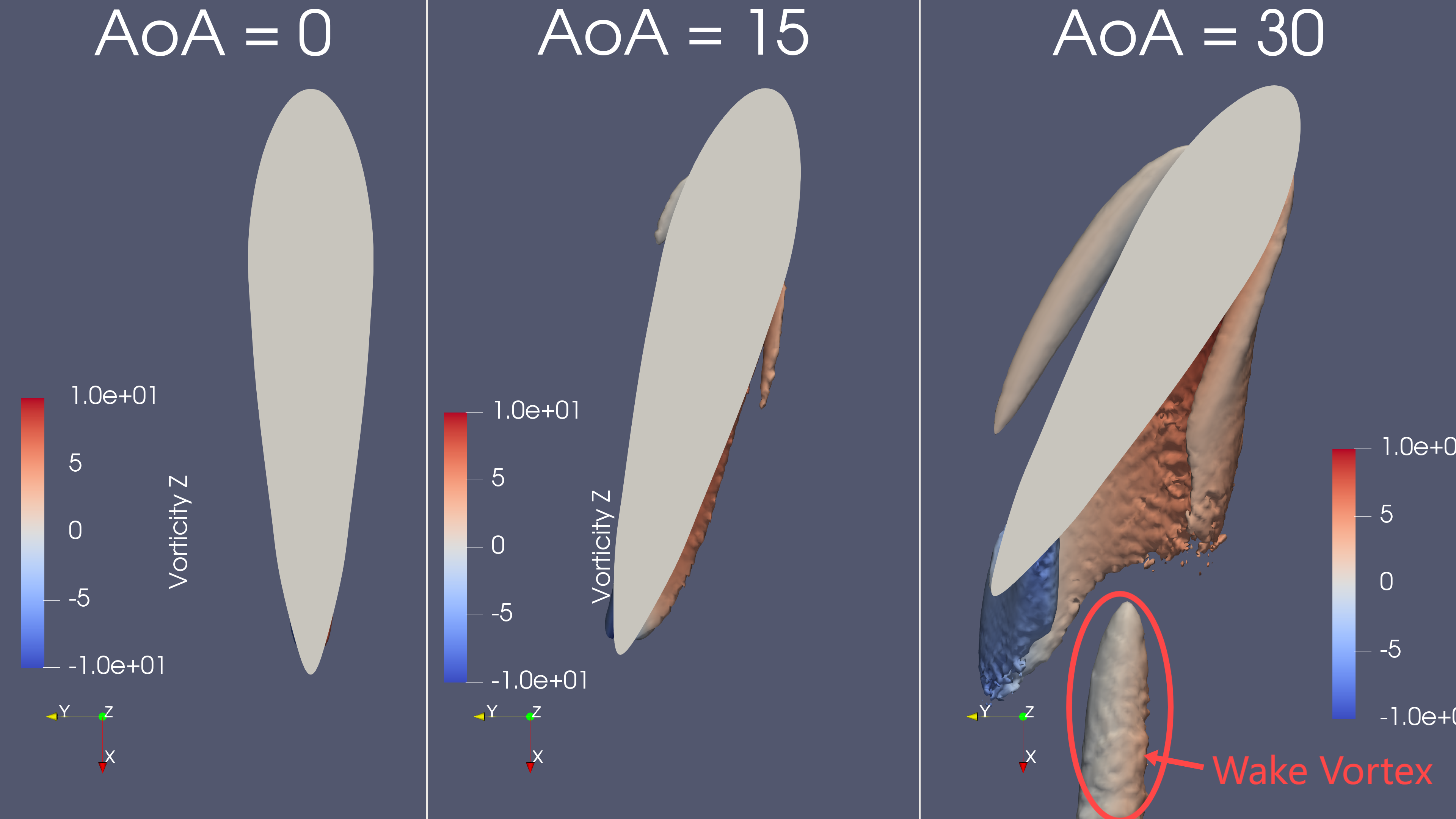}
    \includegraphics[width=1\linewidth]{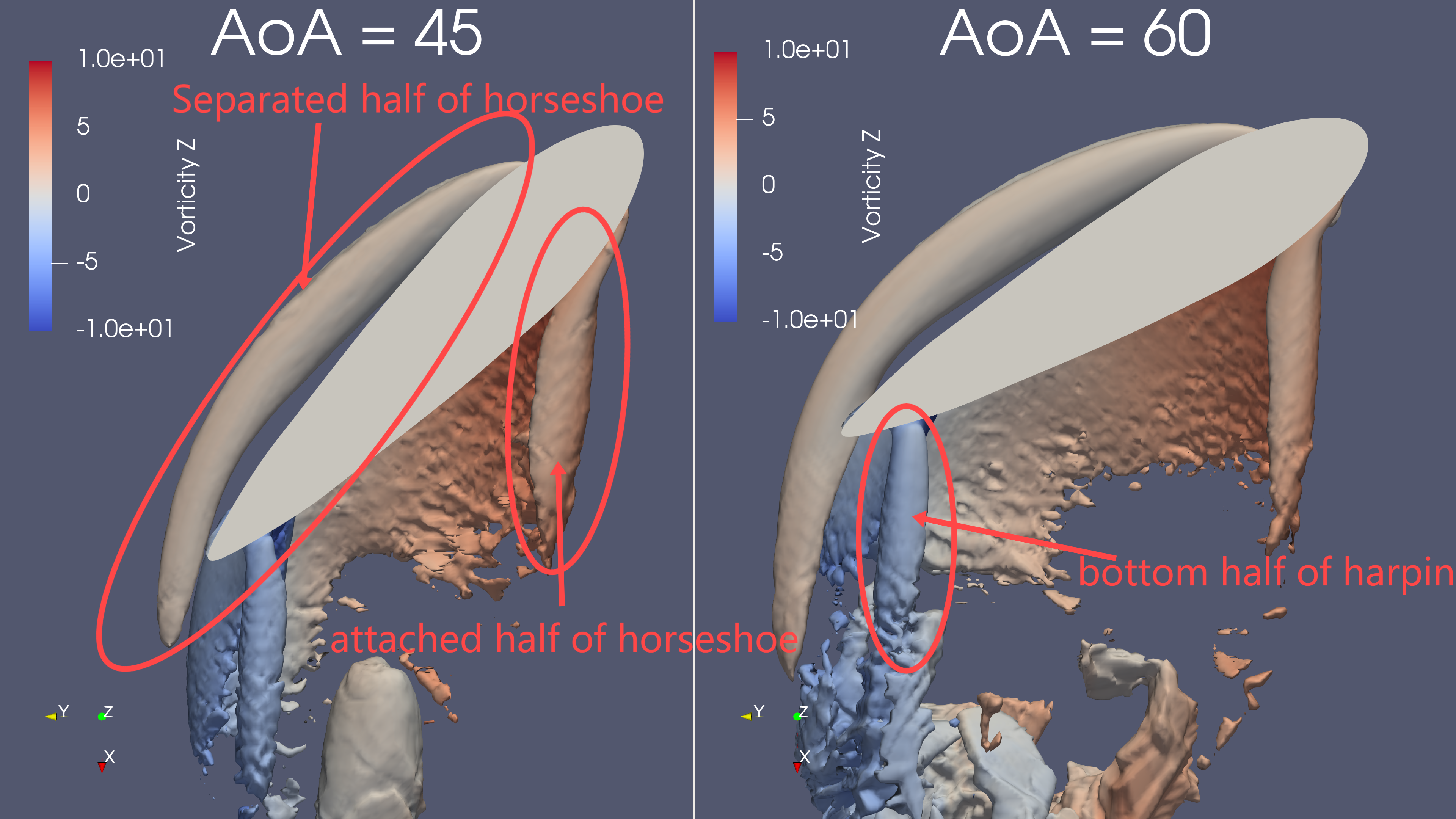}
    \caption{At $\re = 1000$ and $\aoa = 0 - 60\degr$, the isosurface of $Q^* = 7.7 $ is coloured by vorticity in the Z direction, viewed from the bottom. The two halves of the horseshoe vortices can be seen clearly from the bottom angle; both develop significantly with the AoA. The attached half of the horseshoe merges with the arch vortex. The bottom half of the hairpin vortex appears at $\aoa \geq 30$, while extending with $\aoa$. }
    \label{fig:q_vortex_from_bot}
\end{figure}

\begin{figure}
	\centering
	\includegraphics[width=1\linewidth]{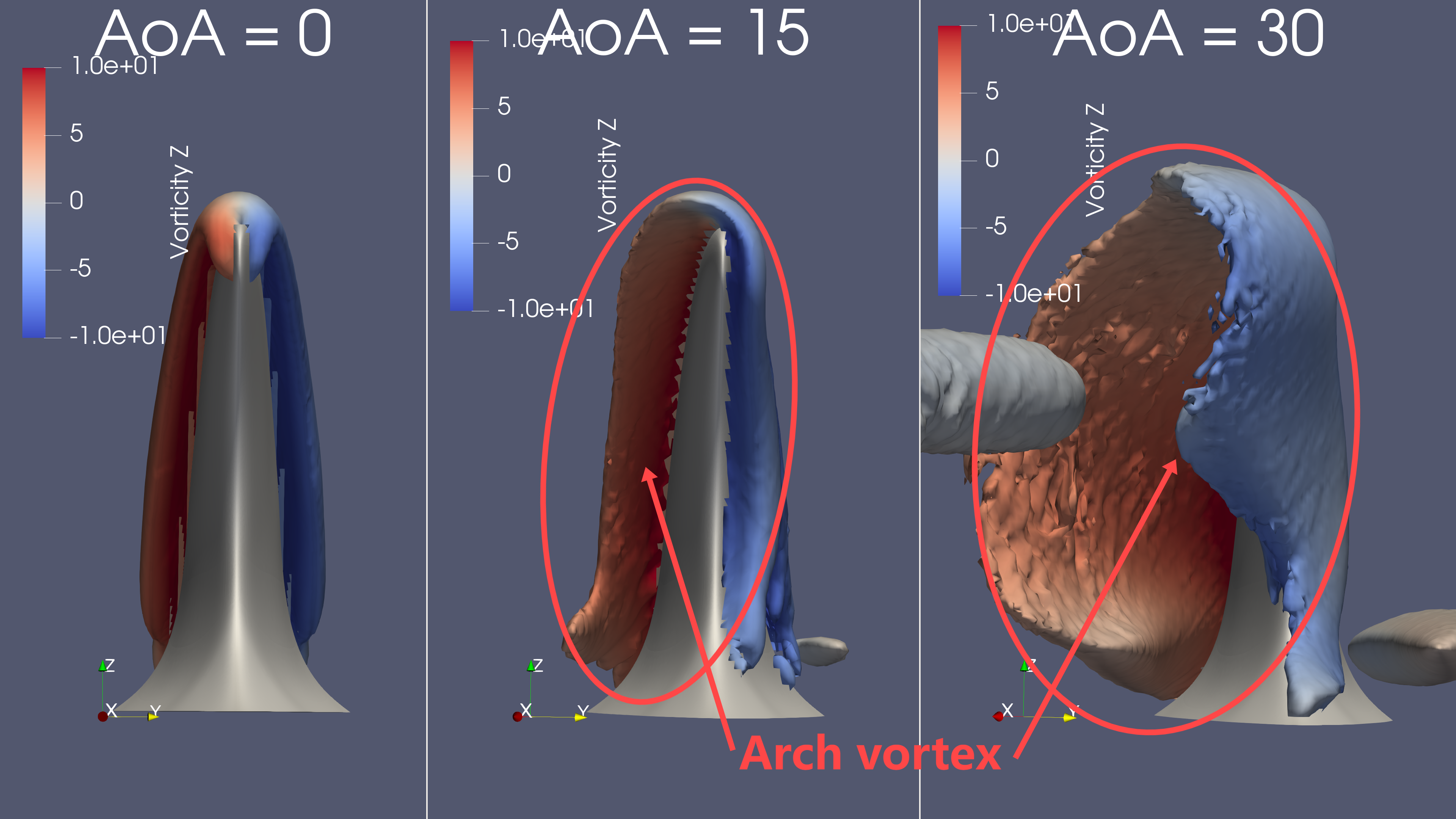}
	\includegraphics[width=1\linewidth]{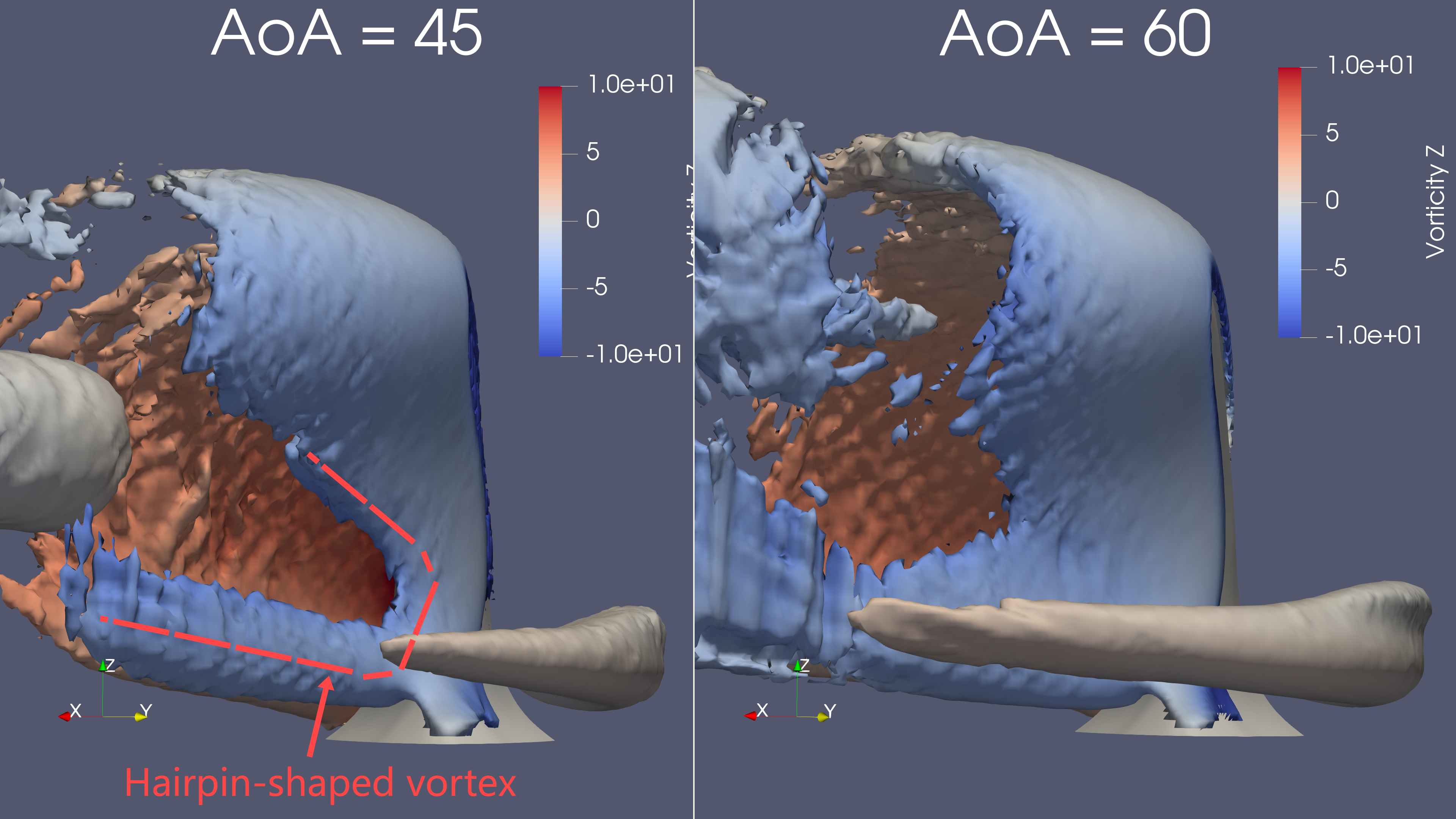}
	\caption{At $\re = 1000$ and $\aoa = 0 - 60\degr$, the isosurface of $Q^* = 7.7 $ is coloured by vorticity in the Z direction, viewed from the back and along the fin's symmetric plane at each angle of attack.
    As $\aoa$ increases, the enveloping arch vortex tilts up and extends further from the downstream side of the dorsal fin, eventually merging with the hairpin-shaped vortex at $\aoa = 60\degr$.
    }%
	\label{fig:q_vortex_arch}
\end{figure}

The Q-criterion is used to visualise the 3-D vortical structures, with a dimensionless value of $ Q^*=Q c^2 U^2 = 7.7 $, as seen in \Cref{fig:aoas_Re_1000_Q_top,fig:q_vortex_from_bot,fig:q_vortex_arch,fig:hair_pin_like_vortex}.
In the following discussion, the naming of vortex types follows conventions established in studies involving flow past wall-mounted objects \cite{Saha2013,Zhang2017,Rastan2017,Martinez-Sanchez2023}.
As depicted in \Cref{fig:aoas_Re_1000_Q_top,fig:q_vortex_arch}, the flow structure becomes significantly more complicated as $\aoa$ increases from $0\degr$ to $60\degr$.
Generally, the geometry of the dolphin fin avoids the complex flow structures typically seen at $\aoa \leq 30\degr$, whereas a square cylinder in a similar Reynolds number range produces much more intricate flow patterns \cite{Saha2013}.

At $\aoa=0\degr$, the flow pattern is symmetric, with no significant flow separation. In this configuration, the dorsal fin is aligned with the incoming flow, and its streamlined geometry effectively minimises the formation of surrounding vortices. This is in stark contrast to the flow patterns observed at higher angles of attack ($\aoa \geq 30\degr$) and the behaviour of bluff bodies at low Reynolds numbers \cite{Saha2013,Zhang2017}.

At $\aoa \geq 30\degr$, abundant flow structures emerge, which correspond to the plateauing of the lift coefficient ($C_L$) observed in \Cref{fig:dragcoe}b.
Two halves of the \textit{horseshoe} vortices form on either side of the dorsal fin at $\aoa \geq 30\degr$, as shown in \Cref{fig:aoas_Re_1000_Q_top,fig:q_vortex_from_bot}. One half detaches from the dorsal fin, resembling the horseshoe vortex structure seen around a square cylinder \cite{Saha2013}, while the other half remains attached to the fin, merging with the arch vortex, as depicted in \Cref{fig:q_vortex_arch}.

The separated half of the horseshoe vortex can be observed on the right side of the dorsal fin when viewed from above, as shown in \Cref{fig:aoas_Re_1000_Q_top}.
As $\aoa$ increases, the separated half-horseshoe extends around the right side of the dorsal fin and into the wake.
At $\aoa = 15\degr$, this vortex is barely noticeable, but at $\aoa = 30\degr$, its length increases more than fourfold.
From $\aoa = 30\degr$ to $45\degr$, its size doubles, its length increasing by another 1.5 times. 
By the time $\aoa$ reaches $60\degr$, the half-horseshoe vortex shows only minor size expansion and little further lengthening.
Thus, the separated half of the horseshoe vortex undergoes significant changes only within the range of $\aoa = 15\degr - 45\degr$.

Similarly, the attached half-horseshoe vortex can be seen on the left side of the dorsal fin when viewed from above, as illustrated in \Cref{fig:aoas_Re_1000_Q_top}.
For a clearer view of the attached vortex, a bottom perspective, as shown in \Cref{fig:q_vortex_from_bot}, better reveals its extension with increasing AoA.
At $\aoa = 15\degr$, the attached vortex is barely visible, but by $\aoa = 30\degr$, both its length and size double. 
However, from $\aoa = 45\degr$ and beyond, the attached vortex maintains its size and length.
Therefore, the attached half of the horseshoe vortex exhibits significant changes only in the range of $\aoa = 15\degr - 30\degr$, which does not align exactly with the changes observed in the separated vortex.

A layer of the arch vortex forms closely above the dorsal fin, as shown in \Cref{fig:q_vortex_arch}.
As $\aoa$ increases, the arch vortex tilts upward and extends further downstream from the dorsal fin, eventually merging with the hairpin-shaped vortex.
At $\aoa = 0\degr - 15\degr$, the arch vortex exhibits only a slight tilt, while at $\aoa = 30\degr$, it extends approximately three times further downstream.
At $\aoa = 45\degr$, the arch vortex on the back of the fin extends about twice as much, connecting with the base vortex to form a hairpin-shaped structure.
The shape of this arch vortex is similar to those observed in the flow around a square cylinder \cite{Saha2013,Martinez-Sanchez2023}.

The "hairpin-shaped vortex" is particularly noteworthy.
In this case, the U-shaped vortex emerges immediately from the crescent-shaped curve on the back of the dolphin dorsal fin, which differs from square-cylinder scenarios where hairpin vortices form further downstream due to a different mechanism. A view of the hairpin vortex from another angle is shown in \Cref{fig:hair_pin_like_vortex}.

The wake vortex is also an interesting aspect to discuss.
It forms at $\aoa = 30\degr - 60\degr$, as seen in \Cref{fig:aoas_Re_1000_Q_top,fig:q_vortex_from_bot}, and its size increases with $\aoa$.
The single wake vortex at $\aoa = 30\degr$ extends directly downstream, which contrasts sharply with the more complex flow structures observed in the wake of a finite square cylinder at similarly low $\re$ \cite{Saha2013,Rastan2017}.

\FloatBarrier
\subsection{Pressure and velocity distribution}
This section examines the variation of pressure and velocity distribution with changing $\aoa$.

\begin{figure}
	\centering
	\includegraphics[width=1\linewidth]{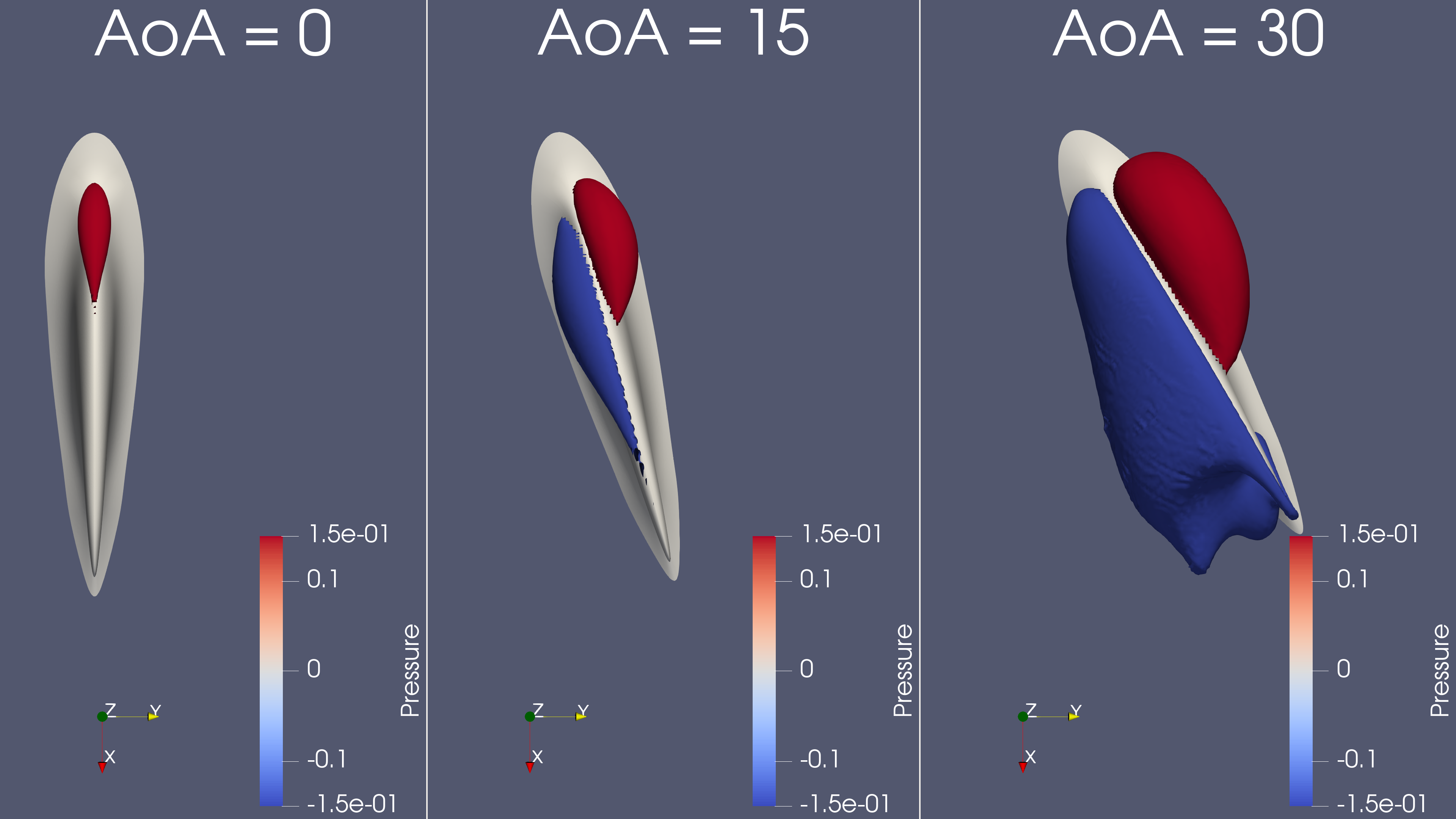}
	\includegraphics[width=1\linewidth]{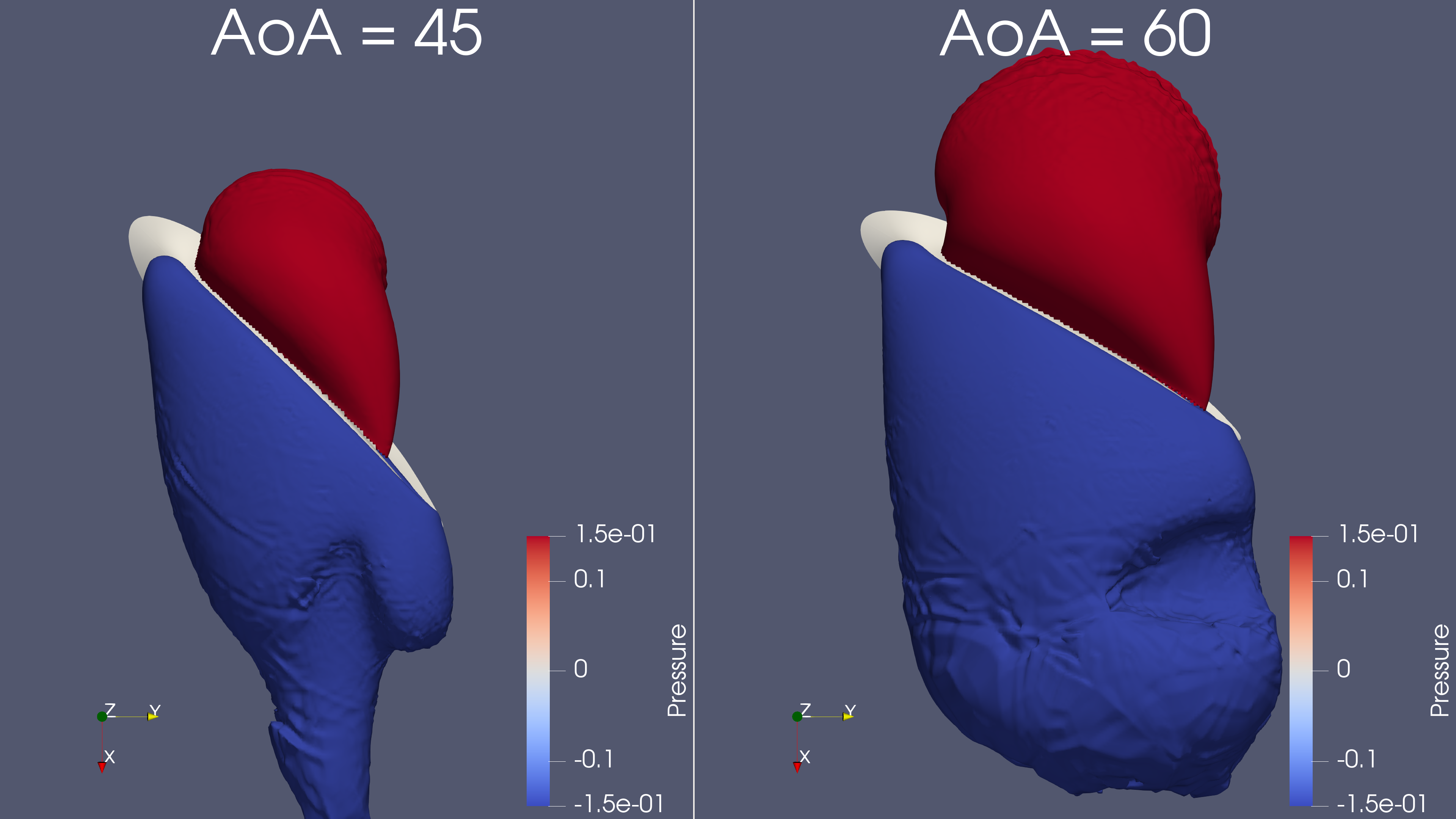}
	\caption{At $\re = 1000$ and $\aoa = 0 - 60$, isosurface of $p^* = 0.15 $ and $-0.15 $ coloured by pressure magnitude, viewed from top.}%
	\label{fig:aoas_re_1000_P_top}
\end{figure}

The pressure distribution provides insight into the force distribution on the dorsal fin.
Here, the pressure is non-dimensionalised as $p^* = p/(\rho U^2) $.
The pressure distribution is shown in \Cref{fig:aoas_re_1000_P_top}.
As the dorsal fin rotates to higher $\aoa$, the pressure distribution changes significantly.
At $\aoa = 0\degr$, positive pressure is observed on the surface facing the incoming flow.
At $\aoa = 15\degr$, positive pressure is still present on the upstream side, while negative pressure appears on the downstream side. 
The magnitudes of both positive and negative pressure increase as $\aoa$ continues to rise.

\begin{figure}
	\centering
	\includegraphics[width=1\linewidth]{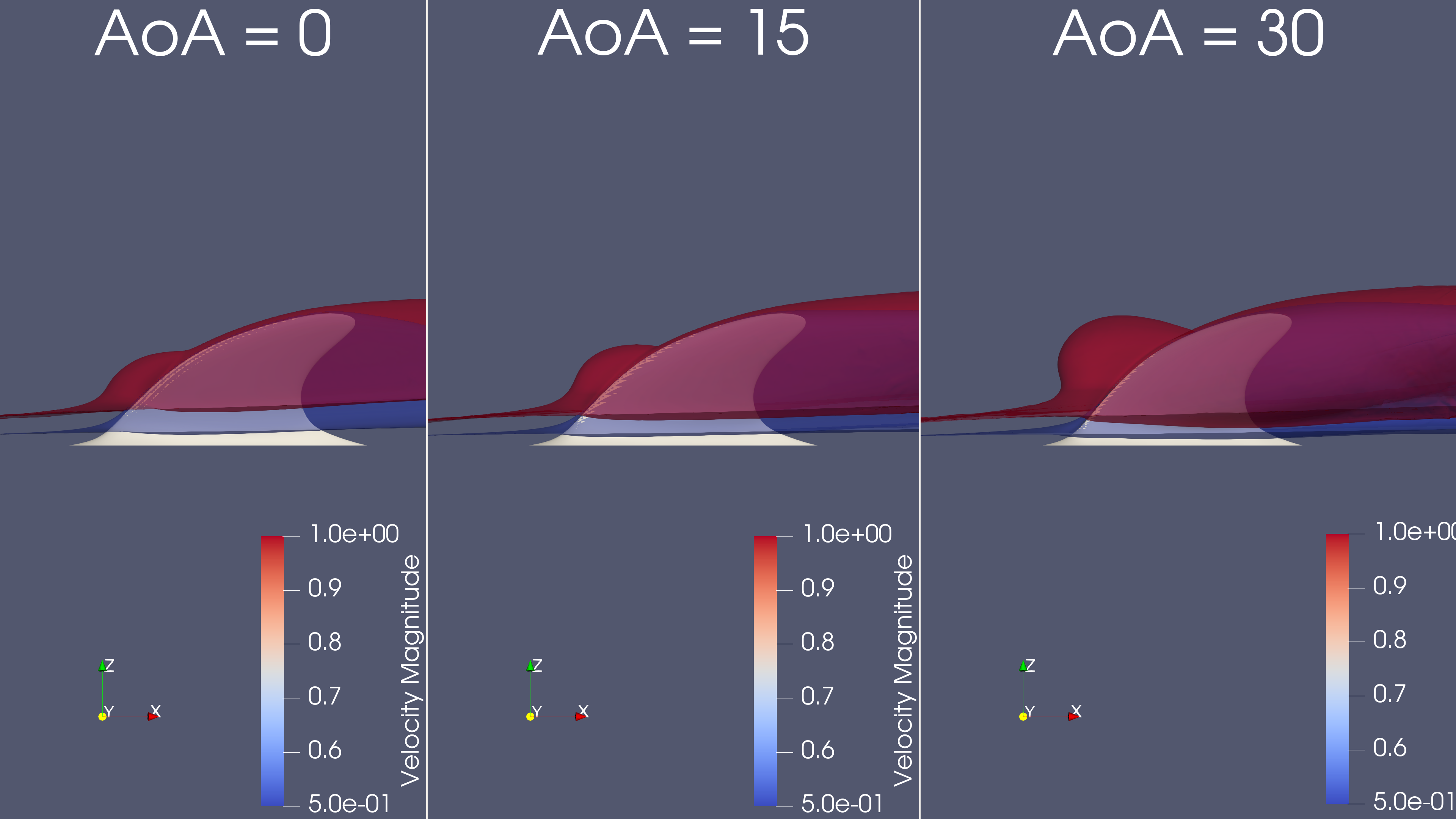}
	\includegraphics[width=1\linewidth]{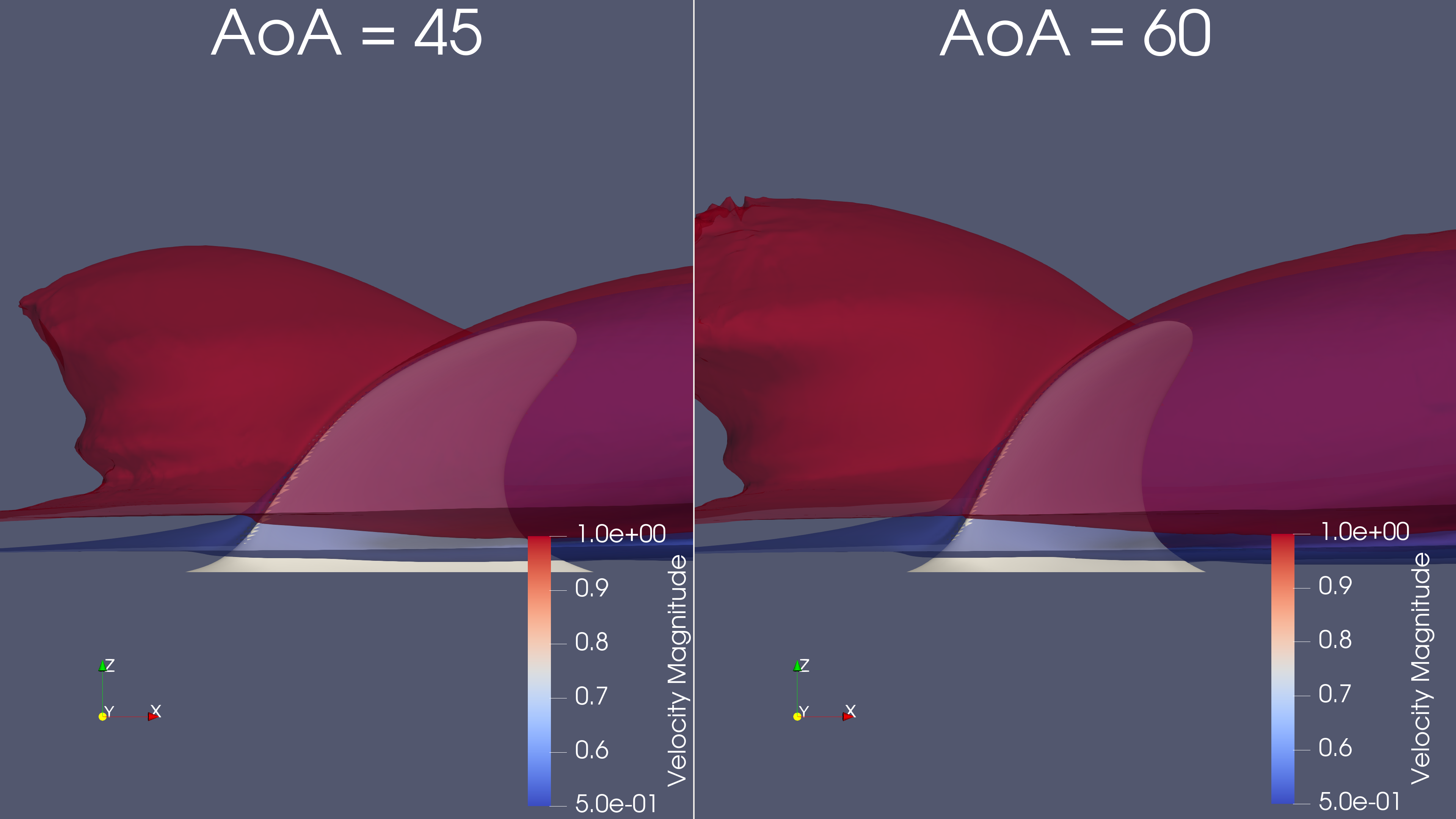}
	\caption{At $\re = 1000$ and $\aoa = 0 - 60\degr$, isosurface of $p^* = 0.15 $ and $-0.15 $ coloured by velocity magnitude, viewed from the side.}
	\label{fig:aoas_re_1000_Vel_top}
\end{figure}

Isosurfaces of local velocity magnitude $V = \sqrt{u^2 + v^2 + w^2}$ are useful for analysing the velocity aspects of the flow structure and serve as an indicator of local kinetic energy.
Two isosurfaces of velocity magnitude $V = 0.5$ and $V = 1.0$ are plotted in \Cref{fig:aoas_re_1000_Vel_top}.
As the AoA increases, the isosurface of $V = 0.5$ shows little variation, while the isosurface of $V = 1.0$ forms a noticeable bump on the upstream side. 
This bump grows significantly as the $\aoa$ increases from $0\degr$ to $60\degr$.

\section{Conclusions}

In this study, we presented a numerical investigation of the flow past a wall-mounted dolphin dorsal fin at angles of attack (AoAs) ranging from $0\degr$ to $60\degr$ and low Reynolds numbers between 691 and 2000. These simulations offer insights into the fundamental flow structures that arise when a dolphin glides through oblique currents or performs turning manoeuvres in water.
Both drag and lift increase significantly with AoA, while drag shows only a slight decrease with increasing Reynolds number. The vortex structures, pressure, and velocity magnitude distributions vary substantially with AoA, and these variations generally align with changes in the lift coefficient.
At $\aoa = 0\degr$, the streamlined shape of the dorsal fin suppresses almost all vortex formation. At $\aoa \geq 30\degr$, several distinct vortical structures emerge, which are generally simpler than those observed around a square cylinder at similar Reynolds numbers. Notably, two half-horseshoe vortices are observed: one upstream, which separates from the fin, and another downstream, which remains attached to the fin. A hairpin-shaped vortex appears near the trailing edge, while an arch vortex envelops the downstream fin surface, connecting the attached half-horseshoe vortex with the hairpin vortex. A clear single wake vortex is found at $\aoa = 30\degr$ and $45\degr$.
The pressure magnitude generally increases with AoA, and the velocity magnitude on the upstream side of the fin also rises as $\aoa$ increases.

This study is limited by the assumption of a rigid fin, as it does not account for fluid-structure interactions of the flexible dorsal fin. Future work will incorporate these interactions to better understand the dorsal fin's function. Additionally, we plan to explore higher Reynolds numbers in future studies.

\vspace{6pt}

\authorcontributions{Zhonglu Lin = Z.L., An-Kang Gao = A.G., Yu Zhang = Y.Z. ``Conceptualization, Z.L. and A.G.; methodology, Z.L. and A.G.; software, Z.L. and A.G.; validation, Z.L.; formal analysis, A.G.; investigation, Z.L. and A.G.; resources, Y.Z.; data curation, Z.L.; writing---original draft preparation, Z.L.; writing---review and editing, A.G.; visualization, Z.L. and A.G.; supervision, A.G. and Y.Z.; project administration, Y.Z.; funding acquisition, A.G. and Y.Z. All authors have read and agreed to the published version of the manuscript.'', please turn to the  \href{http://img.mdpi.org/data/contributor-role-instruction.pdf}{CRediT taxonomy} for the term explanation. Authorship must be limited to those who have contributed substantially to the work~reported.}

\funding{
    We sincerely appreciate the technical assistance regarding NEK5000 from Dr. Xing-Liang Lyu.
    This work was funded by
	Fujian Provincial Department of Human Resources and Social Security 2023-2024 Visiting Scholar and Research Fellowship Program for High-Level Talents and Outstanding Young Talents.
	China Postdoctoral Science Foundation (Grant No. 2021M691865);
	National Natural Science Foundation of China, 12302320;
	Science and Technology Major Project of Fujian Province in China (Grant No. 2021NZ033016).
	The Special Fund for Marine and Fishery Development of Xiamen (Grant No. 20CZB015HJ01).
	The Water Conservancy Science and Technology Innovation Project of Guangdong (Grant No. 2020-16).
	This work was also financially supported by the National Natural Science Foundation of China (Grant Nos. 12074323; 42106181), the Natural Science Foundation of Fujian Province of China (No. 2022J02003), the China National Postdoctoral Program for Innovative Talents (Grant No. BX2021168) and the Outstanding Postdoctoral Scholarship, State Key Laboratory of Marine Environmental Science at Xiamen University.}

\dataavailability{Data available upon request.}

\conflictsofinterest{The authors declare no conflicts of interest.}

 \FloatBarrier

\appendixtitles{no} %
\appendixstart
\appendix
\section[\appendixname~\thesection]{Validation}
\label{sec:appendix}

\begin{figure}[H] 
	\centering
	\includegraphics[width=1\linewidth]{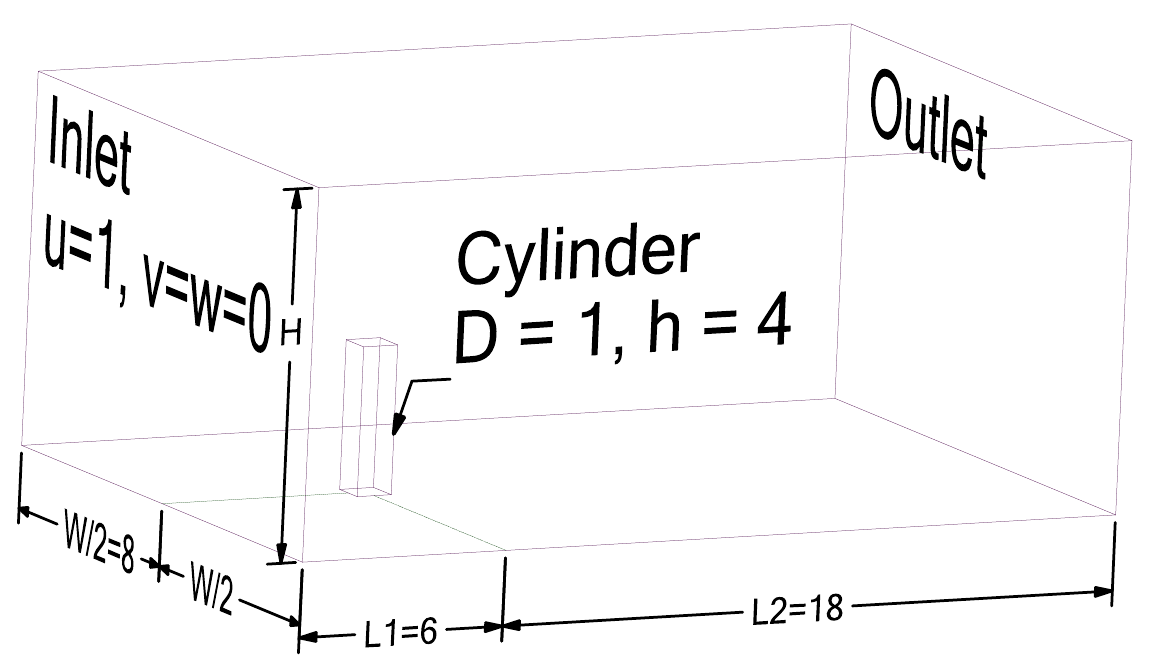}
	\caption{Computational domain of the validation case}
	\label{fig:validation_setup}
\end{figure}

\begin{figure}[H] 
	\centering
	\includegraphics[width=1\linewidth]{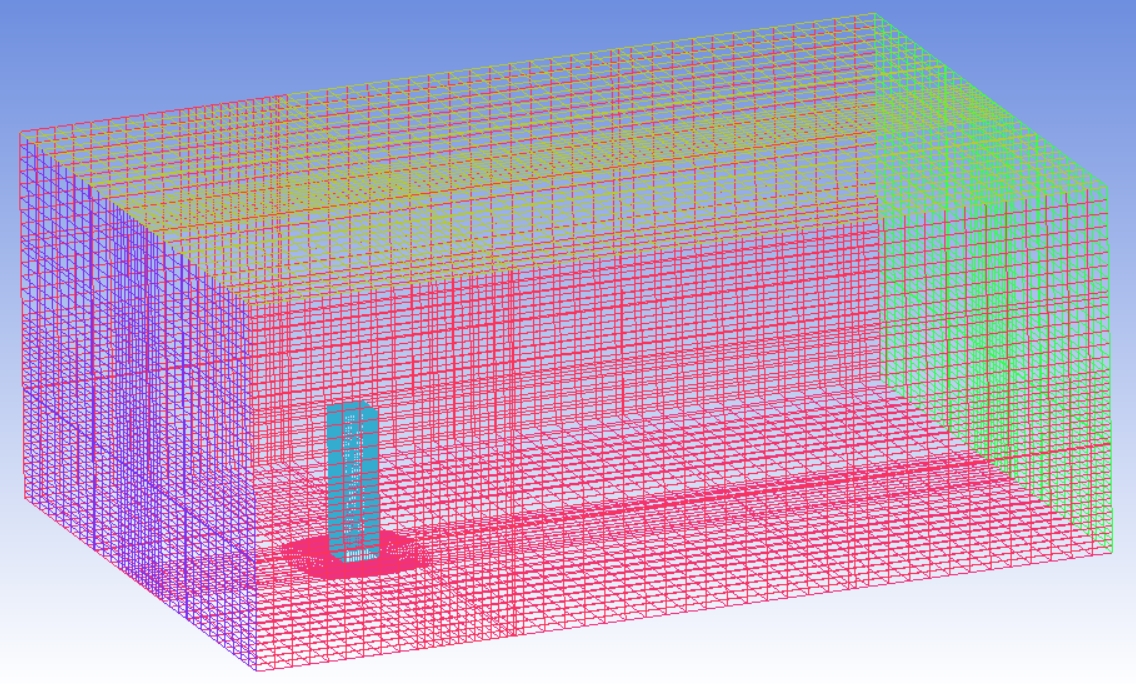}
	\caption{Computational mesh M4 for validation case, O-block is applied to refine mesh around the cylinder}
	\label{fig:validation_mesh}
\end{figure}

To validate the computational setup used in this study with NEK5000, we selected a benchmark case of flow past a finite-length square cylinder, as shown in \Cref{fig:validation_setup,fig:validation_mesh}. 
The problem setup mirrors that used in \cite{Zhang2017,Saha2013}, with a Reynolds number $\re = 250$ and an aspect ratio $AR = 4$, calculated by dividing the characteristic length by the cylinder height. 

The model was validated against the results from \cite{Zhang2017,Saha2013}, as presented in \Cref{tab:validation}. 
The converged mean drag coefficient ($C_d$) from our model, 1.230, closely matched the values reported by \citeauthor{Zhang2017} (1.213) and \citeauthor{Saha2013} (1.23).

\begin{table}[H] 
	\caption{Mesh independence study and validation for a square cylinder of $AR=4$ \label{tab:validation}}
	\newcolumntype{C}{>{\centering\arraybackslash}X}
	\begin{tabularx}{\textwidth}{CCCCCC}
		\toprule 
		\textbf{Name} & \textbf{Grid Resolution} & \textbf{Element Number} $\sci{5}$ & \textbf{$N_x \times N_y \times N_z$} & \textbf{Time Step} & \textbf{Mean Drag Coefficient, Re=250} \\
		\midrule
		M1 & $192 \times 136 \times 124$\textsuperscript{a} & 1.81\textsuperscript{b} & $32 \times 32 \times 64$\textsuperscript{c} & 0.00625 & 1.264  \\ 
		M2 & $200 \times 144 \times 140$ & 2.44 & $40 \times 40 \times 80$ & 0.00500 & 1.243  \\ 
		M3 & $208 \times 152 \times 156$ & 3.24 & $48 \times 48 \times 96$ & 0.00417 & 1.234 \\ 
		M4 & $216 \times 160 \times 172$ & 4.23 & $56 \times 56 \times 112$ & 0.00300 & 1.230 \\ 
		\hline
        Ming's Level-1 & $215 \times 148 \times 125$ & 39.8 & $32 \times 32 \times 77$ & 0.002 & 1.186 \\ 
		Ming's Level-2 & $244 \times 171 \times 148$ & 61.8 & $39 \times 39 \times 93$ & 0.001 & 1.208 \\ 
		Ming's Level-3 & $262 \times 185 \times 170$ & 82.4 & $42 \times 42 \times 104$ & 0.0008 & 1.213 \\ 
		\hline
		Saha's Coarse Level & $152 \times 112 \times 98$ & 16.6 & $26 \times 26 \times 71$ & 0.002 & 1.19 \\ 
		Saha's Middle Level & $177 \times 136 \times 120$ & 28.9 & $34 \times 34 \times 83$ & 0.002 & 1.23 \\ 
		Saha's Refined Level & $204 \times 150 \times 132$ & 40.4 & $42 \times 42 \times 92$ & 0.002 & 1.23 \\ 
		\bottomrule
	\end{tabularx}
	\noindent{\footnotesize{\textsuperscript{a,b,c} For present results obtained by NEK5000 using mesh M1-M4, the grid point number here is for the actual mesh that takes into account the GLL points, calculated as the initial Hex20 element number $\times$ (polynoimal order + 1), which is for the convenience of comparison with other studies based on finite element method.}}
\end{table}

\FloatBarrier

\section[\appendixname~\thesection]{Additional Data}

\begin{figure}
    \centering
    \includegraphics[width=1\linewidth]{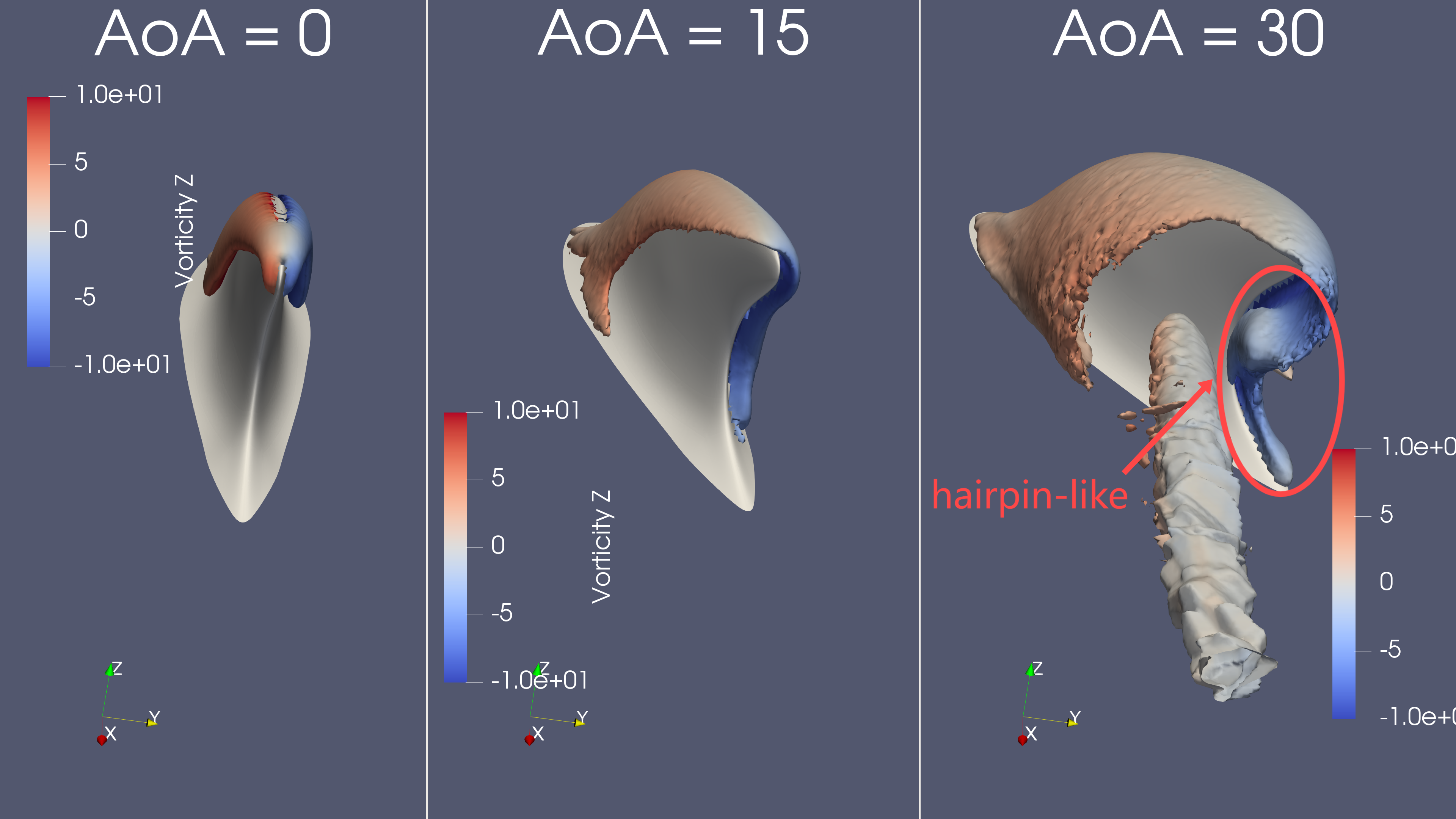}
    \caption{At $\re = 1000$ and $\aoa = 0 - 60\degr$, isosurface of $Q^* = 7.7 $ coloured by $z$-vorticity, viewed from back.
    A hairpin-like vortex can be observed at both $\aoa = 30\degr, 45\degr$. Also, a clear wake vortex can be found in the downstream region.
    }
    \label{fig:hair_pin_like_vortex}
\end{figure}

\begin{figure}
	\centering
	\includegraphics[width=1\linewidth]{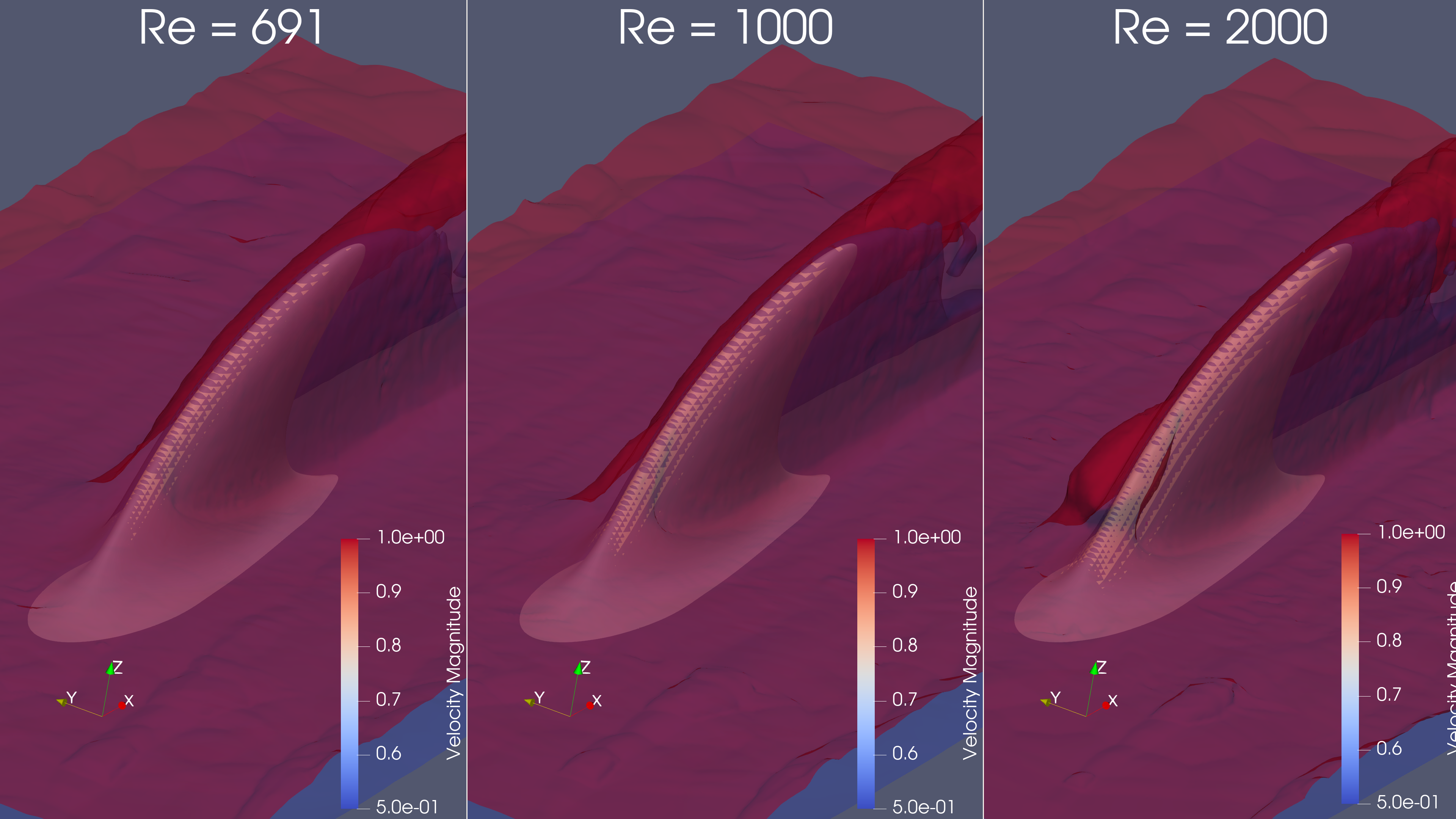}
	\caption{At $\aoa = 0\degr$ and $\re = 691 - 2000$, the Perspective view for isosurfaces of velocity magnitude is shown. The red and blue surfaces stand for where local velocity magnitude $V$ is equal to 1 and 0.5 times of the inlet velocity magnitude $U$.
 }
	\label{fig:aoa0_U}
\end{figure}

\begin{figure}
	\centering
	\includegraphics[width=1\linewidth]{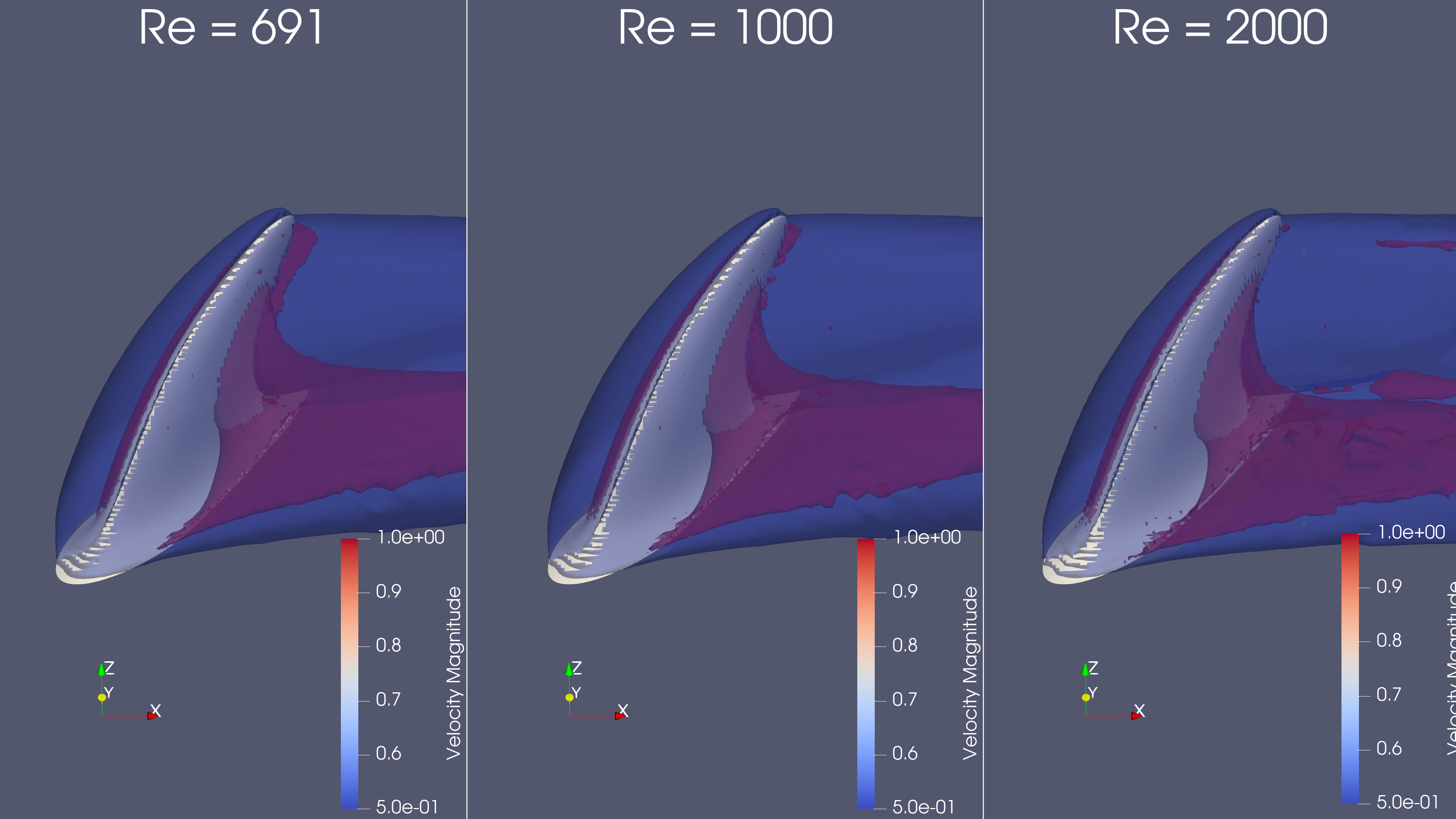}
	\caption{At $\aoa = 60\degr$ and $\re = 691 - 2000$, perspective view for isosurfaces of velocity magnitude is shown.}
	\label{fig:aoa60_U}
\end{figure}

\FloatBarrier
\begin{adjustwidth}{-\extralength}{0cm}

\reftitle{References}

\bibliography{library}

\PublishersNote{}
\end{adjustwidth}
\end{document}